\documentclass[12pt, draftclsnofoot, onecolumn]{IEEEtran}
\usepackage{graphicx}
\usepackage{caption}
\ifCLASSOPTIONcompsoc
  \usepackage[caption=false,font=normalsize,labelfont=sf,textfont=sf]{subfig}
\else
  \usepackage[caption=false,font=footnotesize]{subfig}
\fi
\usepackage{indentfirst}
\usepackage{amsmath}
\usepackage{amssymb}
\usepackage{times}
\usepackage{mathtools}
\usepackage{psfrag}
\usepackage{cite}
\usepackage{lastpage}
\usepackage{fancyhdr}
\usepackage{color}
 \usepackage{amsthm}
\usepackage{bigints}
\newtheorem{theorem}{Theorem}
\newtheorem{Proposition}{Proposition}
\newtheorem{Lemma}{Lemma}
\newtheorem{Remark}{Remark}
\newtheorem{Corollary}{Corollary}
\newtheorem{lemma}[Lemma]{$\mathbf{Lemma}$}

\begin{document}
\title{On the Performance of  Network NOMA in Uplink CoMP {\color{black}S}ystems: A Stochastic Geometry Approach}
\author{Yanshi Sun, Zhiguo Ding, \IEEEmembership{Senior Member, IEEE}, Xuchu Dai, and {\color{black}Octavia A. Dobre, \IEEEmembership{Senior Member, IEEE}}
\thanks{Y. Sun and X. Dai are with the Key Laboratory of Wireless-Optical Communications, Chinese Academy of Sciences, School of Information Science and Technology, University of Science and Technology of China,
No. 96 Jinzhai Road, Hefei, Anhui Province, 230026, P. R. China. (email: sys@mail.ustc.edu.cn, daixc@ustc.edu.cn).

Z. Ding is with the School of Electrical and Electronic Engineering, the University of Manchester, Manchester M13 9PL, U.K. (email:zhiguo.ding@manchester.ac.uk).

{\color{black}O. A. Dobre is with the Faculty of Engineering and Applied
Science, Memorial University, St. John’s, NL A1B 3X5, Canada (e-mail: odobre@mun.ca))}.
}}
\maketitle
\begin{abstract}
To {\color{black}improve} the system throughput, this paper {\color{black}proposes a} network non-orthogonal multiple access (N-NOMA) technique {\color{black}for} the uplink coordinated multi-point transmission (CoMP).
In the considered scenario, multiple base stations collaborate with each other to serve a {\color{black}single} user, referred to as {\color{black}the} CoMP user, {\color{black}which is the same as {\color{black}for} conventional CoMP}. {\color{black}However, unlike conventional CoMP},  each base station {\color{black}in N-NOMA} opportunistically serves an extra user, referred to as {\color{black}the} NOMA user, {\color{black}while serving the CoMP user at the same bandwidth}.
{\color{black}The CoMP user is typically located far from the base stations}, whereas {\color{black}users close to the base stations are scheduled as NOMA users}.
Hence, the channel conditions of the two kind of users are very distinctive, {\color{black}which facilitates the implementation of NOMA}.
Compared to the conventional orthogonal multiple access based CoMP scheme, {\color{black}where multiple base stations serve a single CoMP user only}, the proposed N-NOMA scheme can {\color{black}support larger connectivity by serving {\color{black}the} extra NOMA users, and improve the spectral efficiency by avoiding the CoMP user solely occupying the spectrum}.
A stochastic {\color{black}geometry} approach is applied to model the considered {\color{black}N-NOMA} scenario as {\color{black}a {\color{black}Poisson} cluster process}, based on which closed-form analytical expressions for outage probabilities and ergodic rates are obtained.
Numerical results are presented to show the accuracy of the {\color{black}analytical results} and also demonstrate the superior performance of the proposed {\color{black}N-NOMA} scheme.
\end{abstract}
\section{Introduction}
\IEEEPARstart{R}ecently, {\color{black}non-orthogonal multiple access (NOMA) has attracted significant research attentions in both academia and industry community, not only due to its superior spectral efficiency but also because of its compatibility with other advanced communication} techniques\cite{saito2013non,ding2017NOMAsurvey,ding2017application,ovtavia2017powerdomain}. The key idea of NOMA is to serve more than one user in each orthogonal channel resource block, e.g., a time slot, a frequency channel, a spreading code, or an orthogonal spatial degree of freedom. NOMA has been recognized as a key enabling multiple access technique for the fifth generation (5G) mobile networks. For example, downlink NOMA has been adopted by 3GPP-LTE systems as multiuser superposition transmission (MUST)\cite{GPP2015MUST}. Moreover, NOMA has been recently proposed for the forthcoming digital TV standard (ATSC 3.0), where it is referred to as layered division multiplexing (LDM)\cite{zhang2016LDM}.

\subsection{Related Literature}
The concept of NOMA was initially proposed in \cite{saito2013non} for future wireless netwroks.
The performance of NOMA with randomly deployed users in a downlink scenario was investigated in \cite{ding2014performance}, which {\color{black}shows} that NOMA can outperform conventional {\color{black}orthogonal multiple access (OMA)} in terms of outage performance and ergodic sum rates.
To enhance the performance of {\color{black}users} with weak channel conditions, cooperative NOMA was proposed in \cite{ding2015cooperative} by treating {\color{black}users} with strong channel condition as {\color{black}relays}.
The user fairness of NOMA was studied in \cite{timotheou2015fairness}, and in \cite{ding2014impact}, the authors characterized the impact of user pairing on the performance of two {\color{black}types of} NOMA systems, i.e., NOMA with fixed power allocation (F-NOMA) and cognitive-radio-inspired NOMA (CR-NOMA).
In \cite{xu2017optimal,octavia2017resource,zeng2018energy}, resource allocation for NOMA was investigated. In \cite{hina2017uplinkNOMAPCP} and \cite{ZekunZhang2017densenetwork}, the authors studied the performance of uplink NOMA by considering inter-cell interference.


{\color{black}Note that, \cite{ding2014performance,ding2015cooperative,timotheou2015fairness,ding2014impact,xu2017optimal,octavia2017resource,hina2017uplinkNOMAPCP,ZekunZhang2017densenetwork} focus on single-carrier NOMA, where the principle of NOMA is implemented on a single resource block.  Besides single-carrier NOMA, there are also NOMA schemes termed multi-carrier NOMA, where the principle of NOMA is implemented on multiple resource blocks, e.g, sparse code multiple access (SCMA) \cite{nikopour2013sparse} and pattern-division multiple access (PDMA) \cite{dai2014successive}. Different from single-carrier NOMA which mainly exploits power-domain for multiplexing, multi-carrier NOMA can exploit code-domain across multiple resource blocks to further improve system performance.}

Furthermore, it has been shown that NOMA can be {\color{black}combined} with many other advanced communication techniques. For example, the application of NOMA in millimeter-wave (mmWave) communications was investigated in  \cite{ding2017randommmwave,ding2017nomammwave}. The combination of NOMA and multiple-input multiple-output (MIMO) techniques was studied in  \cite{ding2016general,zeng2017capacity,choi2015minimum,octavia2017MIMONOMA}. Moreover, NOMA was also applied {\color{black}to}  relay systems \cite{yang2017relayselection}, as well as {\color{black}to} wireless caching \cite{Ding2017NOMAcaching}.
\subsection{Motivations and Contributions}

Coordinated multipoint (CoMP) has been recognized as an important enhancement for LTE-A\cite{jungnickel2009coordinated,sawahashi2010coordinated,irmer2011coordinated}. CoMP was mainly proposed to improve cell-edge users' data rates and hence improve the cell coverage.
{\color{black}Conventional CoMP schemes  \cite{venturino2010coordinated,dahrouj2010coordinated} in the literature are based on OMA, where multiple base stations cooperatively serve a single user.}
A drawback of these OMA based schemes is that, once a channel resource block is occupied by a cell-edge user, it cannot be accessed by other users.
Thus, the spectral efficiency becomes worse as the number of cell-edge users increases.
Note that, the channel {\color{black}connection} of a cell-edge user to a base station {\color{black}can be} much worse than that of a user {\color{black}close} to the corresponding base station, due to large scale path loss.
Inspired by this observation, network NOMA (N-NOMA) schemes were proposed in the downlink CoMP systems \cite{choi2014non,sys2017nnomafeasibility}.
{\color{black}The key idea of N-NOMA is to schedule additional users close to the base stations, and} allow cell-edge users and near users to be served simultaneously at the same channel resource block, by combining {\color{black}CoMP which harvests spatial degrees of freedom} with {\color{black}NOMA which improves the spectral efficiency}.
More specifically, in \cite{choi2014non}, {\color{black}the} Alamouti code was applied to improve the cell-edge user's reception reliability, while in \cite{sys2017nnomafeasibility}, {\color{black}distributed} analog beamforming was applied.

The aforementioned N-NOMA schemes in \cite{choi2014non} and \cite{sys2017nnomafeasibility} {\color{black}mainly focus} on the downlink scenario and {\color{black}particular network topologies, e.g., one-dimensional topology is considered in \cite{choi2014non} and equilateral triangle topology in \cite{sys2017nnomafeasibility}}.
Different from \cite{choi2014non} and \cite{sys2017nnomafeasibility}, this paper studies the application of N-NOMA in the uplink CoMP scenario in a {\color{black}general network topology}.
Particularly, a cell-edge user ({\color{black}termed} the CoMP user) is set at the origin, the base stations and their associated near users form a Poisson Cluster process (PCP)\cite{haenggi2012stochastic,chun2015modeling}, where the base stations are treated as parents, and {\color{black}their} associated near users are offsprings.
Each base station chooses a user from its offsprings,  which is referred to as a NOMA user.
The CoMP user and the NOMA users simultaneously transmit messages to the base stations.
After receiving the {\color{black}superimposed} messages, each base station applies {\color{black}successive interference cancellation (SIC)} to first decode its NOMA user's message. If successful, it {\color{black}can} remove the NOMA user's message and then deliver the remaining message {\color{black}as well as the decoded NOMA user's message} to the {\color{black}network controller}, {\color{black}which is connected with all the base stations through wired links}. {\color{black}Finally, the CoMP user's message is decoded at the network controller}.
The contributions of this paper are listed in the following:
\begin{itemize}
  \item The performance of the NOMA users is first investigated.
        The {\color{black}probability density function (pdf)} of the composite channel {\color{black}gain} which consists of  {\color{black}Rayleigh} fading and large scale path loss is first {\color{black}obtained} and approximated by applying the Gaussian-Chebyshev approximation.
        Decoding the NOMA user's message suffers from two kinds of interferences, the interference from the CoMP user and the inter-cell interferences. The Laplace transforms for both kinds of interferences are then {\color{black}applied} to facilitate {\color{black}the} analysis. Closed-form expressions {\color{black}for} the outage probabilities and ergodic rates achieved by the NOMA users are {\color{black}then} obtained.
  \item The study of the CoMP user's performance is divided into two cases.
       \begin{itemize}
         \item In the first case, inspired by cognitive radio (CR),  the NOMA
               users' data rates are adaptively adjusted according to {\color{black}their} instantaneous channel conditions, in order to not degrade the CoMP user's performance {\color{black}compared to OMA based uplink CoMP.}
         \item In the second case, the NOMA {\color{black}users'} date rates are fixed. Note that, only the base stations which can successfully remove {\color{black}their} NOMA {\color{black}users' messages} are allowed to participate {\color{black}in} CoMP, {\color{black}which thins the initial point process. However, in a realization of the point process, the thinning probabilities of different nodes are correlated, since they interfere {\color{black}with} each other. Besides, the thinning probability changes over different realizations of the point process, since the topology is different in different realizations of the point process. The above two facts make the analysis for the outage probability of the CoMP user challenging.}
             To get insight into this case, we turn to study a simplified scheme, namely {\color{black}the} nearest N-NOMA scheme. In {\color{black}the} nearest N-NOMA scheme, only one base station, which is nearest to the CoMP user, is invited to apply NOMA to serve the CoMP user and its NOMA user simultaneously. Closed-form expressions are obtained for outage probabilities achieved by the nearest N-NOMA scheme.
       \end{itemize}
  \item All the {\color{black}analytical results} are validated by computer simulations.  The comparison between the proposed N-NOMA scheme and OMA is illustrated, and it is concluded that the proposed N-NOMA scheme significantly improves the spectral efficiency compared to OMA.
\end{itemize}

\subsection{Organization}
The rest of this paper is organized as follows. Section II illustrates the system model. Section III analyzes the  performance of the proposed N-NOMA scheme. Section IV provides numerical results to demonstrate the performance of the proposed N-NOMA scheme and also verify the accuracy of the developed analytical results. Section V concludes the paper. Finally, Appendixes collect the proofs of the obtained analytical results.

\section{System Model}
\subsection{System Model}
\begin{figure}[!t]
\centering
\includegraphics[width=3.5in]{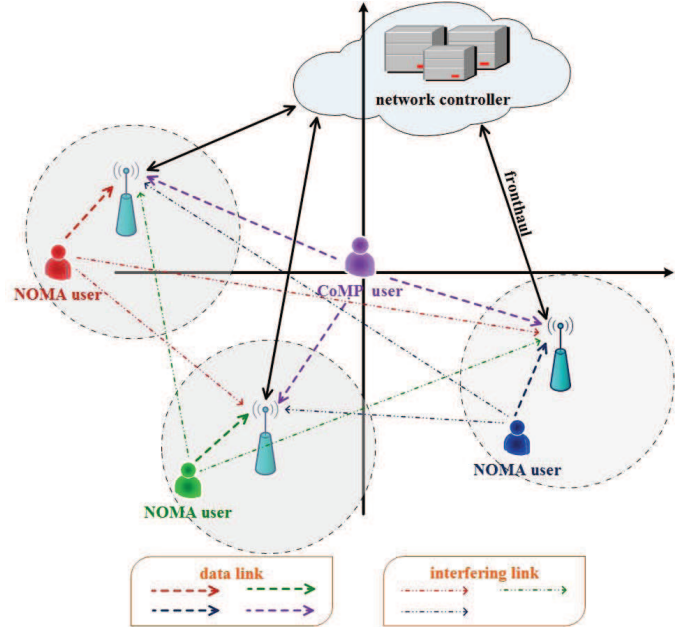}
\caption{An illustration of the system model.}
\label{system_model}
\end{figure}
Consider {\color{black}an} N-NOMA communication scenario as explained in the following. {\color{black}As illustrated in Fig. \ref{system_model},} there are multiple  base stations {\color{black}which} collaborate with each other to help a {\color{black}single} CoMP user, denoted by $\text{U}_0$, which is similar to conventional CoMP. In addition, each base station will opportunistically serve one extra user at the same time and frequency channel allocated to $\text{U}_0$.   The extra user served by each base station is referred to as a NOMA user in this paper.

Without loss of generality, the CoMP user is placed at the origin. The locations of the base stations and the NOMA users are modeled as a PCP. In particular, assume that the locations of the base stations are denoted by $x_i$ and modeled as a homogeneous Poisson point process (HPPP), denoted by $\Phi_c$, with density $\lambda_{c}$, i.e., $x_i\in \Phi_c$. Each base station is the parent node of a cluster covering a disk whose radius is denoted by  $\mathcal{R}_c$. To implement NOMA transmission, the base station in cluster $i$, denoted by $\text{BS}_i$,  invites $K$ users, denoted by $\text{U}_{i,k}$, {\color{black}$k\in\{1,\ldots,K\}$}, to participate in NOMA transmission. These users associated with the same base station are viewed as offspring nodes. The locations of these users, denoted by $y_{i,k}$, {\color{black}$k\in\{1,\ldots,K\}$}, are uniformly distributed in the disk with $\text{BS}_i$ located at its origin. To simplify the notation, the locations of the cluster users are conditioned on the locations of their cluster heads. {\color{black}As such, the distance from a cluster user to its cluster head is $||y_{i,k}||$, and the distance from $\text{U}_{i,k}$ to $\text{BS}_j$ is $||y_{i,k}+x_i-x_j||$.}

{\color{black}Among $K$ users associated with $\text{BS}_i$}, {\color{black}$\text{U}_{i,k_i^*}$} is selected {\color{black}for N-NOMA transmission}, {\color{black}and the user selection criterion  is to select the user with }the largest  composite channel gain as follows:
\begin{align}\label{user selection}
{\color{black}k_i^*} = \underset{k\in \{1, \cdots, K\}}{\arg\max} \frac{{\color{black}|h_{\text{BS}_i,\text{U}_{i,k}}|^2}}{L\left(||y_{i,k}||\right)}.
\end{align}
where  {\color{black}$h_{\text{BS}_i,\text{U}_{j,k}}$} denotes the Rayleigh fading channel coefficient between $\text{BS}_i$ and $\text{U}_{j,k}$,  and $L\left(||y_{j,k}||\right)$ denotes the path loss. Particularly, the following path loss model is used, $\frac{\eta}{L\left(||y_{j,k}||\right)}$, where $L\left(||y_{j,k}||\right)=||y_{j,k}||^\alpha$, {\color{black}$\eta=\frac{c^2}{16\pi^2f_c^2}$ denotes the parameter relevant to carrier frequency $f_c$ ($c$ is light speed)}, and  $\alpha$ denotes the path loss exponent.
It is worth pointing out that, throughout the paper, all base stations use the same user selection {\color{black}criterion as shown in} (1).

{\color{black}In this paper, it is assumed that all nodes are equipped with a single antenna. This assumption is applicable in many scenarios in {\color{black}forthcoming} 5G networks. For example, in applications that combine Internet-of-things (IoT) with cellular networks, randomly deployed low-cost IoT {\color{black}access points} are more likely to have a single antenna.}

It is also worth pointing out that,  in this paper, the distance between the CoMP user and a base station is {\color{black}typically} much {\color{black}larger} than {\color{black}the distance of a NOMA user}. Under this circumstance, the channel conditions will be distinctive enough {\color{black}to facilitate the implementation of NOMA}.
\subsection{Description of N-NOMA}
For uplink N-NOMA, the CoMP user, {\color{black}denoted by} $\text{U}_0$,  transmits its signal to all the base stations.  In addition, the user with {\color{black}the largest composite channel gain} in cluster $i$ is invited to send its  information to $\text{BS}_i$. Therefore, $\text{BS}_i$ receives the following:
\begin{align}\label{model1}
y_{\text{BS}_i} =& \underset{\text{CoMP user's signal}}{\underbrace{\frac{\sqrt{\eta P_0}{\color{black}h_{\text{BS}_i,\text{U}_0}}}{\sqrt{L\left(||x_i||\right)}} s_0}}  +  \underset{\text{NOMA user's signal}}{\underbrace{\frac{\sqrt{\eta P}{\color{black}h_{\text{BS}_i,\text{U}_{i,k_i^*}}}}{\sqrt{L\left(||{\color{black}y_{i,k_i^*}}||\right)}} {\color{black}s_{i,k_i^*}}}}
\\ \nonumber
&+  \underset{\text{Inter-cell interference}}{\underbrace{\sum_{x_j\in \Phi_c\backslash x_i} \frac{\sqrt{\eta P}{\color{black}h_{\text{BS}_i,\text{U}_{j,k_j^*}}}}{\sqrt{L\left(||{\color{black}y_{j,k_j^*}}+x_j-x_i||\right)}} {\color{black}s_{j,k_j^*}}}} +n_i,
\end{align}
where  $P_0$ denotes the {\color{black}transmission} power of the CoMP user, $P$ denotes the NOMA users' {\color{black}transmission} power, ${\color{black}s_{i,k_i^*}}$ denotes the message sent by ${\color{black}\text{U}_{i,k_i^*}}$, $s_0$ is the message sent by $\text{U}_0$, {\color{black}$h_{\text{BS}_i,\text{U}_0}$} is defined similarly to ${\color{black}h_{\text{BS}_i,\text{U}_{i,k}}}$, and $n_i$ denotes the additive noise, which is modeled as a circular symmetric complex {\color{black}Gaussian} random variable, i.e., $n_i \sim \mathcal{CN}(0,\sigma^2)$, where $\sigma^2$ is the noise power.

Among the three types of information in \eqref{model1}, $\text{BS}_i$ will first decode its NOMA user's information, since this user in cluster $i$ is very likely to be much  closer to $\text{BS}_i$ compared to the CoMP user. If successfully, $\text{BS}_i$ will then subtract the NOMA user's signal from its observation, and pass the remaining signal as
well as the decoded NOMA user's message to the {\color{black}network controller}. {\color{black}Finally, the CoMP user's message is decoded at the network controller. The corresponding {\color{black}signal-to-interference-plus-noises (SINRs)} for decoding these signals are provided in the following.

\subsubsection{Decoding the NOMA user's signal} As discussed before, the  NOMA users' signals are decoded first, by treating {\color{black}the CoMP user's} signals as interference.  In cluster $i$, $\text{BS}_i$ decodes its NOMA user's information
 with the following SINR ratio:
\begin{align}
 {\color{black}\text{SINR}_{k_i^*}^{\text{BS}_i}} = \frac{\frac{|{\color{black}h_{\text{BS}_i,\text{U}_{i,k_i^*}}}|^2}{L\left(||{\color{black}y_{i,k_i^*}}||\right)}}{ I_{\text{CoMP}}^U +I_{\text{inter}}^U
 +\frac{1}{\rho}},
\end{align}
where $\rho=\frac{\eta P}{\sigma^2}$, $I_{\text{CoMP}}^U  = \frac{\phi |{\color{black}h_{\text{BS}_i,\text{U}_0}}|^2}{ {L\left(||{\color{black}x_i}||\right)}}$ denotes the interference from the CoMP user {\color{black}with $\phi=\frac{P_0}{P}$},
and the inter-cluster interference is given by
\begin{align}
I_{\text{inter}}^U=\sum_{x_j\in \Phi_c\backslash x_i} \frac{|{\color{black}h_{\text{BS}_i,\text{U}_{j,k_j^*}}}|^2}{ {L\left(||{\color{black}y_{j,k_j^*}}+x_j-x_i||\right)}}.
\end{align}
The corresponding achievable data rate of ${\color{black}\text{U}_{i,k_i^*}}$ is given by
\begin{align}
{\color{black}R_{i,k_i^*}}=\log_2(1+{\color{black}\text{SINR}_{k_i^*}^{\text{BS}_i}}).
\end{align}

\subsubsection{Decoding the CoMP user's signal}
Provided that $\text{SINR}_{k_i^*}^{\text{BS}_i}\geq \epsilon_i$, where $\epsilon_{i}=2^{\tilde{R}_{i,{\color{black}k_i^*}}}-1$ and $\tilde{R}_{i,{\color{black}k_i^*}}$ is the date rate {\color{black}of} ${\color{black}\text{U}_{i,k_i^*}}$,
$\text{BS}_i$ can decode the message sent by ${\color{black}\text{U}_{i,k_i^*}}$ and hence remove this message from its observation successfully.
{\color{black}
Note that not all base stations can  decode  the messages from their {\color{black}corresponding} NOMA users, which further thins the original HPPP, $\Phi_c$.
Denote by $\bar{\Phi}_c$ the new point process including {\color{black}these} qualified base stations.
Moreover, inviting all the base stations in the plane to participate in CoMP {\color{black}could result in prohibitive system complexity due to the coordination among the qualified base stations}.
{\color{black}Because of the aforementioned} two reasons, it is assumed that only those base stations which can remove their {\color{black}corresponding} NOMA users' information and are within disk $\mathcal{D}$ (which is centered at the origin and with radius $\mathcal{R}_{\mathcal{D}}$),  are allowed to participate in CoMP.
The modified {\color{black}observation expression} after removing the NOMA user's message at $\text{BS}_i$ can be expressed as follows:
\begin{align}
\bar{y}_{\text{BS}_i}=& {\frac{\sqrt{\eta P_0}{\color{black}h_{\text{BS}_i,\text{U}_0}}}{\sqrt{L\left(||x_i||\right)}} s_0}+  {{\sum_{x_j\in \Phi_c\backslash x_i} \frac{\sqrt{\eta P}{\color{black}h_{\text{BS}_i,\text{U}_{j,k_j^*}}}}{\sqrt{L\left(||{\color{black}y_{j,k_j^*}}+x_j-x_i||\right)}} {\color{black}s_{j,k_j^*}}}} +n_i.
\end{align}
After each base station subtracts its NOMA user's signal from its observation, {\color{black}it forwards} its modified observation as well as its NOMA user's message to the network controller.

At the network controller, each forwarded $\bar{y}_{\text{BS}_i}$ can be further {\color{black}processed} by removing the interference term $s_{j,k^*_j}$, if $\text{BS}_j \in \bar{\Phi}_c \cap \mathcal{D}$, since the network controller has collected those NOMA users' messages. Then $\bar{y}_{\text{BS}_i}$ {\color{black}can be further rewritten as follows:}
\begin{align}
\tilde{y}_{\text{BS}_i}=& {\frac{\sqrt{\eta P_0}{\color{black}h_{\text{BS}_i,\text{U}_0}}}{\sqrt{L\left(||x_i||\right)}} s_0}+  {{\sum_{x_j\in \Phi_c\backslash (\bar{\Phi}_c \cap \mathcal{D}) } \frac{\sqrt{\eta P}{\color{black}h_{\text{BS}_i,\text{U}_{j,k_j^*}}}}{\sqrt{L\left(||{\color{black}y_{j,k_j^*}}+x_j-x_i||\right)}} {\color{black}s_{j,k_j^*}}}} +n_i.
\end{align}

For each $\tilde{y}_{\text{BS}_i}$, the SINR to decode the CoMP user's message {\color{black}is given by}
\begin{align}
 \text{SINR}_{0}^{\text{BS}_i} = \frac{I_{\text{CoMP}}^U}{\tilde{I}_{\text{inter}}^U
 +\frac{1}{\rho}},
\end{align}
where
\begin{align}
\tilde{I}_{\text{inter}}^U=\sum_{x_j\in \Phi_c\backslash (\bar{\Phi}_c \cap \mathcal{D})} \frac{|{\color{black}h_{\text{BS}_i,\text{U}_{j,k_j^*}}}|^2}{ {L\left(||{\color{black}y_{j,k_j^*}}+x_j-x_i||\right)}}.
\end{align}
{\color{black}Finally, the network controller selects the base station which has the largest $\text{SINR}_{0}^{\text{BS}_i}$ to decode the CoMP user's message
 and the corresponding achievable data rate of the CoMP user at the network controller is given by}
\begin{align}
R_0=\max_{\substack{\text{BS}_i \\ x_i\in \Phi_c\backslash (\bar{\Phi}_c \cap \mathcal{D})}}\log_2(1+ \text{SINR}_{0}^{\text{BS}_i}).
\end{align}
}

\section{Performance Analysis for Uplink N-NOMA}
This section studies the outage probability and the ergodic rate (or average data rate) achieved by {\color{black}the NOMA and CoMP users}.
The outage probability is the probability of the event that the data rate supported by the instantaneous channel realizations is less than the targeted data rate.
{\color{black}Accordingly}, the outage probability is an important performance evaluation metric of {\color{black}the} quality of service (QoS) in delay-sensitive communication scenarios, where the information sent by the transmitter is at a fixed rate.
The average data rate can be used for the case {\color{black}when} the transmitted data rates are determined adaptively, according to the users' channel conditions. This case corresponds to
delay-tolerant communications. To achieve {\color{black}Shannon's} capacity, it is assumed in this paper {\color{black}that} the messages sent by users are independently coded with {\color{black}Gaussian} codebooks.

\subsection{NOMA user}
Consider a typical base station in disk $\mathcal{D}$, say $\text{BS}_i$, we will first study the outage probability for $\text{BS}_i$ to decode its NOMA user's information which is denoted by {\color{black}$\mathrm{P}_{\text{NOMA}}$}. This probability can be expressed as follows:
\begin{align}
{\color{black}\mathrm{P}_{\text{NOMA}}} = \mathrm{P}\left({\color{black}\text{SINR}_{k_i^*}^{\text{BS}_i}}<\epsilon_{i}\right).
\end{align}
Note that $\text{BS}_i$ is a point {\color{black}in} disk $\mathcal{D}$,  {\color{black}belonging to the HPPP $\Phi_c$;} thus it is not hard to conclude that $\text{BS}_i$ is uniformly distributed in $\mathcal{D}$ \cite{haenggi2012stochastic}.

To calculate the probability $ \mathrm{P}\left(\text{SINR}_{{\color{black}k_i^*}}^{\text{BS}_i}<\epsilon_{i} \right)$, the first {\color{black}step} is to determine the distribution of the composite channel gain, $\frac{|{\color{black}h_{\text{BS}_i,\text{U}_{i,k}}}|^2}{L\left(||y_{i,k}||\right)}$. For {\color{black}simplicity of notation}, we define $z_{k,i} \triangleq \frac{|{\color{black}h_{\text{BS}_i,\text{U}_{i,k}}}|^2}{L\left(||y_{i,k}||\right)}$, which means ${\color{black}z_{k_i^*,i}}= \max \{z_{1,i}, \ldots,  z_{K,i}\}$.
For an unordered channel gain, i.e., $z_{k,i}$, we can apply the {\color{black}Gaussian}-Chebyshev approximation as in \cite{ding2014performance} to {\color{black}obtain} the following lemma:
\begin{lemma}
The cumulative distribution function (CDF) for an unordered composite channel gain, i.e., $z_{k,i}$ is given by:
\begin{eqnarray}\label{cdf_ch}
F_{z_{k,i}}(z) \approx   \sum^{N}_{n=1}w_n  \left(1-e^{-c_nz}\right),
\end{eqnarray}
and the corresponding pdf is given by:
\begin{align}\label{pdf_ch}
f_{z_{k,i}}(z) \approx  \sum^{N}_{n=1}w_nc_n  e^{-c_nz},
\end{align}
where $w_n=  \frac{\pi}{2N}\sqrt{1-\theta_n^2}
\left(\theta_n+1\right)$, $N$ is the {\color{black}Gaussian}-Chebyshev parameter, $c_n=\left(\frac{\mathcal{R}_c}{2}\theta_n+\frac{\mathcal{R}_c}{2}\right)^\alpha$,   and $\theta_n =\cos \left( \frac{2n-1}{2N} \pi\right)$.
\end{lemma}
As can be seen in (\ref{cdf_ch}), the CDF is expressed as the sum of a finite number of exponentials. It is worth pointing out that this representation not only is accurate  with a small $N$, but also can significantly facilitate the derivation, as can be seen later {\color{black}in Appendix B}.

Another {\color{black}step used} to calculate $ \mathrm{P}\left({\color{black}\text{SINR}_{k_i^*}^{\text{BS}_i}}<\epsilon_{i} \right)$ is to determine the Laplace transform for the two interference terms $I_{\text{CoMP}}^U$ and $I_{\text{Inter}}^U$, and the following lemma follows.
\begin{lemma}\label{lemma2}
{\color{black}Given} the coordinate of $\text{BS}_i$ at $x_i$, the Laplace transform for the interference from the CoMP user is given by:
\begin{align}
\mathcal{L}_{I_{\text{CoMP}}^U}(s | x_i)&= \mathcal{E}_{I_{\text{CoMP}}^U } \left\{e^{-s I_{\text{CoMP}}^U} \right\}\\\notag
&=\frac{1}{1+s\phi L\left(||x_i||\right)^{-1}},
\end{align}
and the Laplace transform for the inter-cluster interference is given by:
 \begin{align}
\mathcal{L}_{I_{\text{inter}}^U}(s|x_i) &= \mathcal{E}_{I_{\text{inter}}^U } \left\{e^{-s I_{\text{inter}}^U} \right\}\\\notag
&=\text{exp}\left(-   2\pi\lambda_{c} \frac{s^{\frac{2}{\alpha}}}{\alpha}\text{B}\left(\frac{2}{\alpha}, \frac{\alpha-2}{\alpha}\right) \right),
\end{align}
where $B(\cdot)$ is the beta function.
\end{lemma}
\begin{IEEEproof}
Please refer to Appendix A.
\end{IEEEproof}
Note that in Lemma \ref{lemma2}, the Laplace transform for $I_{\text{CoMP}}^U$ is a function of $||x_i||$. This is {\color{black}due to the fact} that the base station {\color{black}which} is closer to the CoMP user will be more likely to {\color{black}suffer} strong interference from the CoMP user. But the Laplace transform for $I_{\text{Inter}}^U$ {\color{black}is not related} with $x_i$, this {\color{black}is due to} the stationary property of the considered Poisson process.

By applying Lemma 1, Lemma 2 and the property of the order statistics \cite{david2003order}, the following theorem characterizing the outage probability {\color{black}achieved by $\text{BS}_i$ to decode ${\color{black}U_{i,k_i^*}}$'s message} is obtained.
\begin{theorem}
The outage probability achieved by $\text{BS}_i$ to decode ${\color{black}U_{i,k_i^*}}$'s message can be approximated as {\color{black}follows}:
\begin{align}
 {\color{black}\mathrm{P}_{\text{NOMA}}} \approx&1+\sum_{\substack{k_0+\cdots+k_N=K \\ k_0\neq K}}{K \choose k_0, \cdots, k_N} \left(\prod_{n=0}^{N} \tilde{w}_n^{k_n}\right)\times e^{-\mu\epsilon_i
 \frac{1}{\rho}}  \\ \nonumber &\times    {
\text{exp}\left(-   2\pi\lambda_{c} \frac{(\mu\epsilon_i)^{\frac{2}{\alpha}}}{\alpha}\text{B}\left(\frac{2}{\alpha}, \frac{\alpha-2}{\alpha}\right) \right)}\\\notag &\times
\frac{2{\mathcal{R}_{\mathcal{D}}}^\alpha}{\phi\mu\epsilon_i(2+\alpha)}
{}_2\text{F}_1\left(1, 1+\frac{2}{\alpha}; 2+\frac{2}{\alpha}; -\frac{{R_{\mathcal{D}}}^{\alpha}}{\phi\mu\epsilon_i}\right),
\end{align}
where $\tilde{w}_n=-w_n$ for $1\leq n \leq N$ and $\tilde{w}_0=1$,
and $\tilde{c}_n=c_n$ for $1\leq n \leq N$ and $\tilde{c}_0=0$,
and $k_n \geq0$ for $0\leq n \leq N$,
and $\mu= \sum_{n=0,k_n\neq 0}^{N}k_n\tilde{c}_n$,
and ${K \choose k_0, \cdots,k_N}=\frac{K!}{k_0!\cdots k_N!}$,
and ${}_2\text{F}_1(\cdot)$ is the hypergeometric function.
\end{theorem}
\begin{IEEEproof}
Please refer to Appendix B.
\end{IEEEproof}

The ergodic rate achieved by ${\color{black}U_{i,k_i^*}}$ {\color{black}is given by}
\begin{align}
{\color{black}R_{\text{NOMA}}^{ave}}=\mathcal{E}\left\{\log_2(1+\text{SINR}_{{\color{black}k_i^*}}^{\text{BS}_i})\right\}.
\end{align}
By applying {\color{black}Theorem} 1, we can {\color{black}obtain} the following corollary to characterize ${\color{black}R_{\text{NOMA}}^{ave}}$.
\begin{Corollary}
The ergodic rate achieved by ${\color{black}U_{i,k_i^*}}$ can be approximated as:
\begin{align}
{\color{black}R_{\text{NOMA}}^{ave}}\approx&-\sum_{\substack{k_0+\cdots+k_N=K \\ k_0\neq K}}{K \choose k_0, \cdots, k_N} \frac{2{\mathcal{R}_{\mathcal{D}}}^\alpha\prod\limits_{n=0}^{N} \tilde{w}_n^{k_n}}{\ln{2}\phi\mu (2+\alpha)}
\int_0^{\infty}\\\notag
 &\frac{e^{- \frac{\mu}{\rho}x}}{x(1+x)}  \times    {
\text{exp}\left(-   2\pi\lambda_{c} \frac{(\mu x)^{\frac{2}{\alpha}}}{\alpha}\text{B}\left(\frac{2}{\alpha}, \frac{\alpha-2}{\alpha}\right) \right)}\\\notag &\times
{}_2\text{F}_1\left(1, 1+\frac{2}{\alpha}; 2+\frac{2}{\alpha}; -\frac{{R_{\mathcal{D}}}^{\alpha}}{\phi\mu x}\right)\,dx.
\end{align}
\end{Corollary}
\begin{IEEEproof}
Please refer to Appendix C.
\end{IEEEproof}

The use of N-NOMA brings the opportunity for the base stations to serve {\color{black}the corresponding} NOMA users, and hence, improves the system throughput. Thus, it is necessary to quantify the ergodic sum rates of {\color{black}the} NOMA users.  The ergodic sum rate obtained by serving {\color{black}the} NOMA users {\color{black}is} defined as
\begin{align}
{\color{black}R^{sum}_{\text{NOMA}}}=\mathcal{E}\left\{\sum_{\substack{x_i \in \Phi_c \cap \mathcal{D}}}{\color{black}R_{i,k_i^*}}\right\}.
\end{align}
Note that here only the NOMA users whose associated base stations are located in disk $\mathcal{D}$,  are taken into account. A closed-form expression for ${\color{black}R^{sum}_{\text{NOMA}}}$ is {\color{black}provided} as follows.
\begin{Corollary}
The ergodic sum rate {\color{black}for} the NOMA users whose associated base stations are located in disk $\mathcal{D}$ can be calculated as
\begin{align}
{\color{black}R^{sum}_{\text{NOMA}}}= \pi\lambda_c\mathcal{R}_\mathcal{D}^2{\color{black}R^{ave}_{\text{NOMA}}}.
\end{align}
\end{Corollary}
\begin{IEEEproof}
{\color{black}By using the definition of $R_{sum}^{\text{NOMA}}$, {\color{black}this} can be evaluated as:}
\begin{align}
R_{sum}^{\text{NOMA}}&=\mathcal{E}\left\{\sum_{\substack{x_i \in \Phi_c \cap \mathcal{D}}}{\color{black}R_{i,k_i^*}}\right\}\\\notag
&=\mathcal{E}\left\{{\color{black}M}\right\}\cdot\mathcal{E}\left\{{\color{black}R_{i,k_i^*}}\right\}\\\notag
&=\mathcal{E}\left\{{\color{black}M}\right\}\cdot {\color{black}R^{ave}_{\text{NOMA}}},
\end{align}
where ${\color{black}M}$ is the number of points in $\Phi_c \cap \mathcal{D}$. Note that $\Phi_c$ is a HPPP, and thus,  $\mathcal{E}\left\{{\color{black}M}\right\}$ can be expressed as:
\begin{align}
\mathcal{E}\left\{{\color{black}M}\right\}=\pi\lambda_c\mathcal{R}_\mathcal{D}^2.
\end{align}
Therefore, the proof is {\color{black}complete}.
\end{IEEEproof}

\subsection{CoMP user}
\subsubsection{{\color{black}For the case with adaptive} $\tilde{R}_{i,k_i^*}$}
{\color{black} In some scenarios, ensuring the QoS of the CoMP user is of the most importance. In this case, in the considered N-NOMA system, the CoMP user can be treated as {\color{black}a} primary user and the NOMA user {\color{black}can be treated as a} secondary user, as in conventional cognitive networks.
Note that, in the conventional OMA based uplink CoMP systems, as in \cite{sun2018stochastic,sun2017coordinated}, all the base stations in disk $\mathcal{D}$ are employed to serve the CoMP user and are not allowed to serve any NOMA user.
In the considered N-NOMA, if the NOMA users' transmission data rates are adaptively adjusted according to their instantaneous channel conditions, satisfying $\tilde{R}_{i,k_i^*} \leq R_{i,k_i^*}$ can ensure that each base station in disk $\mathcal{D}$ can successfully decode its NOMA user's message.
As such, according to the description in Section II, it is easy to conclude that the CoMP user can achieve the same performance in the proposed N-NOMA as in {\color{black}the} OMA based system.
}
\subsubsection{{\color{black}For the case with fixed} ${\color{black}\tilde{R}_{i,k_i^*}}$}
{\color{black}It is also interesting to study the case where the NOMA users have fixed data rate requirements.}
{\color{black}Note that the outage probability achieved by the CoMP user can be expressed as follows:
\begin{align}
\mathrm{P}_{\text{CoMP}}=
\mathrm{P}\left(\max_{\substack{\text{BS}_i \\ x_i\in \Phi_c\backslash (\bar{\Phi}_c \cap \mathcal{D})}} \text{SINR}_{0}^{\text{BS}_i} < \epsilon_0
\right)
\end{align}
where $\epsilon_0=2^{\tilde{R}_0}-1$, $\tilde{R}_0$ is the date rates of the CoMP user.
}
In this case, only the base stations which can successfully remove their NOMA users' messages are allowed to participate in CoMP, {\color{black}and} the new point process $\bar{\Phi}_c$ including those qualified base stations can be treated as a thinning process of the original point process $\Phi_c$.
The outage probability {\color{black}$P_{\text{NOMA}}$} in Theorem 1 can be treated as a thinning probability of a typical base station. However, this thinning probability is only an average thinning probability of the base stations in disk $\mathcal{D}$, over many realizations of the process.  {\color{black}As} can be expected, the thinning probability of a node will be different in different realizations of the point process. {\color{black}Furthermore},  in each realization, the thinning
probabilities of different nodes are correlated, since they interfere  {\color{black}with} each other. The above observations indicate that the analysis for this case will be very challenging. Thus, we will {\color{black}rely on} simulations {\color{black}for the} performance evaluation for this case, as shown in Section IV. Furthermore, to get insight {\color{black}into} this case, we consider a simplified N-NOMA scheme, {\color{black}referred to as} the nearest N-NOMA scheme, as shown in the following subsection.

\subsection{Nearest N-NOMA scheme}
In the nearest N-NOMA scheme, only one base station, which is nearest to the CoMP user, is invited to apply NOMA to serve the CoMP user and its NOMA user simultaneously.  Similar to the {\color{black}aforementioned  general} N-NOMA scheme, in the nearest N-NOMA  scheme, the nearest base station first {\color{black}tries} to decode its NOMA user's message. {\color{black}If}  successfully, the nearest base station will then subtract the NOMA user’s signal from its observation, and deliver the remaining signal to the {\color{black}network controller}.

Besides the  aforementioned motivation in Section III-B, it is worth pointing out {\color{black}that there is} another motivation to study the nearest N-NOMA scheme. The nearest N-NOMA scheme needs only one base station to serve the CoMP user, and hence, the system overhead is reduced. {\color{black}Accordingly} the nearest N-NOMA scheme  is applicable to scenarios where the system overhead is the bottleneck of the system performance.

In the following, the outage performance of the nearest N-NOMA scheme is analyzed. For notation convenience, in this subsection, the base stations are ordered according to their distances to the CoMP user as follows:
\begin{align}
  {L\left(||{\color{black}x_i}||\right)} \leq     {L\left(||x_j||\right)}
\end{align}
for $i<j$. Thus, the nearest base station is $\text{BS}_1$.
\subsubsection{NOMA user}
Note that according to \cite{haenggi2012stochastic},  $\text{BS}_1$ is not a typical point since it is closest to the CoMP user. Thus, the NOMA user's outage probability for {\color{black}the nearest N-NOMA scheme} will be different from the outage probability obtained in Theorem 1, as highlighted in the following proposition.
\begin{Proposition}
When $\alpha=4$, the NOMA user's outage probability for {\color{black}the} nearest N-NOMA scheme can be approximated as {\color{black}follows:}
\begin{align}
\mathrm{P}_{1}^{\text{nearest}}
&\approx1+\sum_{\substack{k_0+\cdots+k_N=K\\k_0\neq K}}{K \choose k_0, \cdots, k_N} \prod_{n=0,k_n\neq 0}^{N} \tilde{w}_n^{k_n} e^{-k_n\tilde{c}_n\epsilon_1
 \frac{1}{\rho}}  \\ \nonumber &\times
\text{exp}\left(-2\pi\lambda_{c} \frac{(\epsilon_1\mu)^{\frac{2}{\alpha}}}{\alpha}\text{B}\left(\frac{2}{\alpha}, \frac{\alpha-2}{\alpha}\right)\right) \times\\ \notag
& \int_0^{\infty}
\text{exp}\left(\frac{1}{2}\pi\lambda_c\sqrt{\epsilon_1\mu}
\arctan\left(\sqrt{-\frac{1}{2}+\frac{1}{2}\sqrt{1+\frac{16d^4}{\epsilon_1\mu}}}\right)
\right)\\\notag
&\times \frac{1}{1+\phi\epsilon_1\mu d^{-4}}
\times2\lambda_c\pi d e^{ -\lambda_c\pi d^2} \,dd.
\end{align}
\end{Proposition}
\begin{IEEEproof}
Please refer to Appendix D.
\end{IEEEproof}
\subsubsection{CoMP user}
When the NOMA user's rate is fixed, the outage probability achieved by the CoMP user is characterized as shown in the following proposition.
{\color{black}
\begin{Proposition}
When $\alpha=4$, the CoMP user's outage probability for the nearest N-NOMA scheme can be approximated as follows:
\begin{align}\label{P_0_n}
 \mathrm{P}_{0}^{\text{nearest}} \approx&1+ \sum_{\substack{k_0+\cdots+k_N=K \\ k_0\neq K}}{K \choose k_0, \cdots, k_N} \prod_{n=0,k_n\neq 0}^{N} \tilde{w}_n^{k_n} e^{-k_n\tilde{c}_n\epsilon_1
 \frac{1}{\rho}}  \\ \nonumber &\times
\int_0^{\infty}\text{exp}\left(-2\pi\lambda_{c} \frac{\xi(d)^{\frac{2}{\alpha}}}{\alpha}\text{B}\left(\frac{2}{\alpha}, \frac{\alpha-2}{\alpha}\right)\right.\\ \notag
&\left.\quad\quad+\frac{1}{2}\pi\lambda_c\sqrt{\xi(d)}
\arctan\left(\sqrt{-\frac{1}{2}+\frac{1}{2}\sqrt{1+\frac{16d^4}{\xi(d)}}}\right)
\right)\\\notag
&\times \frac{2\pi\lambda_cd^5}{\phi\epsilon_1\mu+d^4}
\text{exp}\left(-\frac{\epsilon_0(\phi\epsilon_1\mu+d^4)}{\phi\rho}-\lambda_c\pi d^2\right)\,dd,
\end{align}
where $\xi(d)=\epsilon_1\mu+{\epsilon_0(\phi\epsilon_1\mu+d^4)}/{\phi}$.
\end{Proposition}
\begin{IEEEproof}
Please refer to Appendix D.
\end{IEEEproof}}
\begin{Remark}
{\color{black}We choose $\alpha=4$ without loss of generality, as the procedure of the analysis is the same.}  {\color{black}In addition,} the geometry property that the distance between the BSs is much larger than that between a NOMA user to its associated BS is {\color{black}used} {\color{black}to facilitate analysis},  as shown in Appendix E.
\end{Remark}

\section{Numerical results}
In this section, computer simulations {\color{black}are performed} to demonstrate the performance of the proposed N-NOMA system and also verify the accuracy of the analytical results. The thermal noise power is set {\color{black}to} $-170$ dBm/Hz, the carrier frequency is $f_c=2\times10^9$ Hz, the transmission bandwidth is $B=10$ MHz, and the transmitter and receiver antenna gains are set {\color{black}to} $1$. The path loss {\color{black}exponent} is set as $\alpha=4$. The path loss parameter is set as $\eta=\frac{c^2}{16\pi^2f_c^2}$, where $c$ is the light speed.
\begin{figure}[!t]
\setlength{\belowcaptionskip}{-1.5em}   
\centering
\includegraphics[width=3.5in]{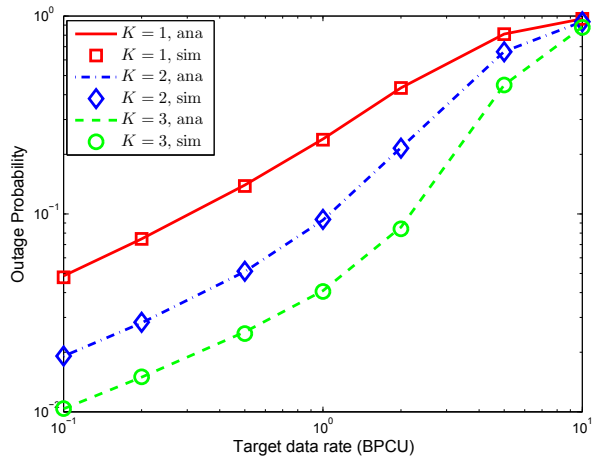}
\caption{NOMA user outage probability versus target data rate. $\mathcal{R}_c=50$ m, $\lambda_c=2\times10^{-5}/m^2$, $P=0.1$ W, $\phi=10$, and the {\color{black}Gaussian}-Chebyshev parameter $N=20$. {\color{black}Notations, sim: simulation results; ana: analytical results.} }
\label{NOMA_outage_1}
\end{figure}

\begin{figure}[!t]
\setlength{\belowcaptionskip}{-1.5em}   
\centering
\includegraphics[width=3.5in]{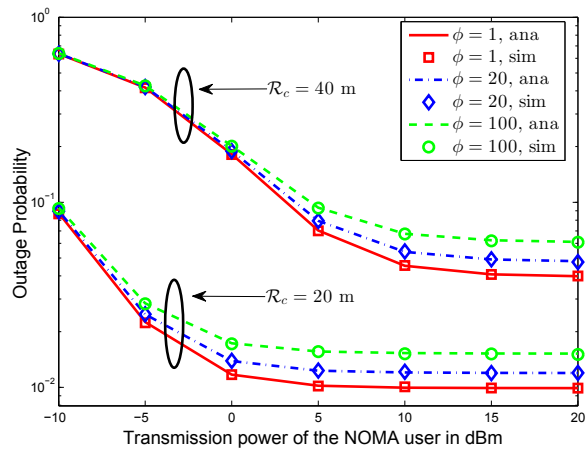}
\caption{NOMA user outage probability versus transmission power. $K=2$, ${\color{black}\tilde{R}_{i,k_i^*}}=1$ bit per channel use (BPCU), $\lambda_c=2\times10^{-5}/m^2$, and the {\color{black}Gaussian}-Chebyshev parameter $N=20$. }
\label{NOMA_outage_2}
\end{figure}

In Fig. \ref{NOMA_outage_1} and Fig. \ref{NOMA_outage_2}, the outage probability of a typical NOMA user is studied.
Fig. \ref{NOMA_outage_1} shows the outage probability as a function of the user {\color{black}target} rate, while Fig. \ref{NOMA_outage_2} {\color{black}shows it} as a function of the NOMA user transmission power.
It should be {\color{black}mentioned} that the average outage probabilities of the NOMA users, {\color{black}whose associated base stations are located in disk $\mathcal{D}$, are considered in both figures}. The radius of $\mathcal{D}$ is set as $\mathcal{R}_\mathcal{D}=500$ m.  As shown in {\color{black}Figs. \ref{NOMA_outage_1} and \ref{NOMA_outage_2}}, computer simulations perfectly  match the {\color{black}theoretical} results, which demonstrates the accuracy of the developed analysis.
From Fig. \ref{NOMA_outage_1}, it is {\color{black}can be seen} that as the number of users in a cluster increases, the outage probability of the NOMA user {\color{black}decreases}. This is because the NOMA user is chosen from the $K$ users in the cluster according to the channel conditions.
From Fig. \ref{NOMA_outage_2}, it is shown that as the transmission power of  the CoMP user increases, the outage probability of a NOMA user {\color{black}also increases}.  The reason is that when decoding the NOMA user's message, the CoMP user's message is treated as noise.
From Fig. \ref{NOMA_outage_2} it is also {\color{black}seen} that the outage probability of a NOMA user increases with the radius of a cluster.

\begin{figure*}[!t]
\setlength{\belowcaptionskip}{-1.5em}   
\centering
\subfloat[Erdogic rate]{\includegraphics[width=3.2in]{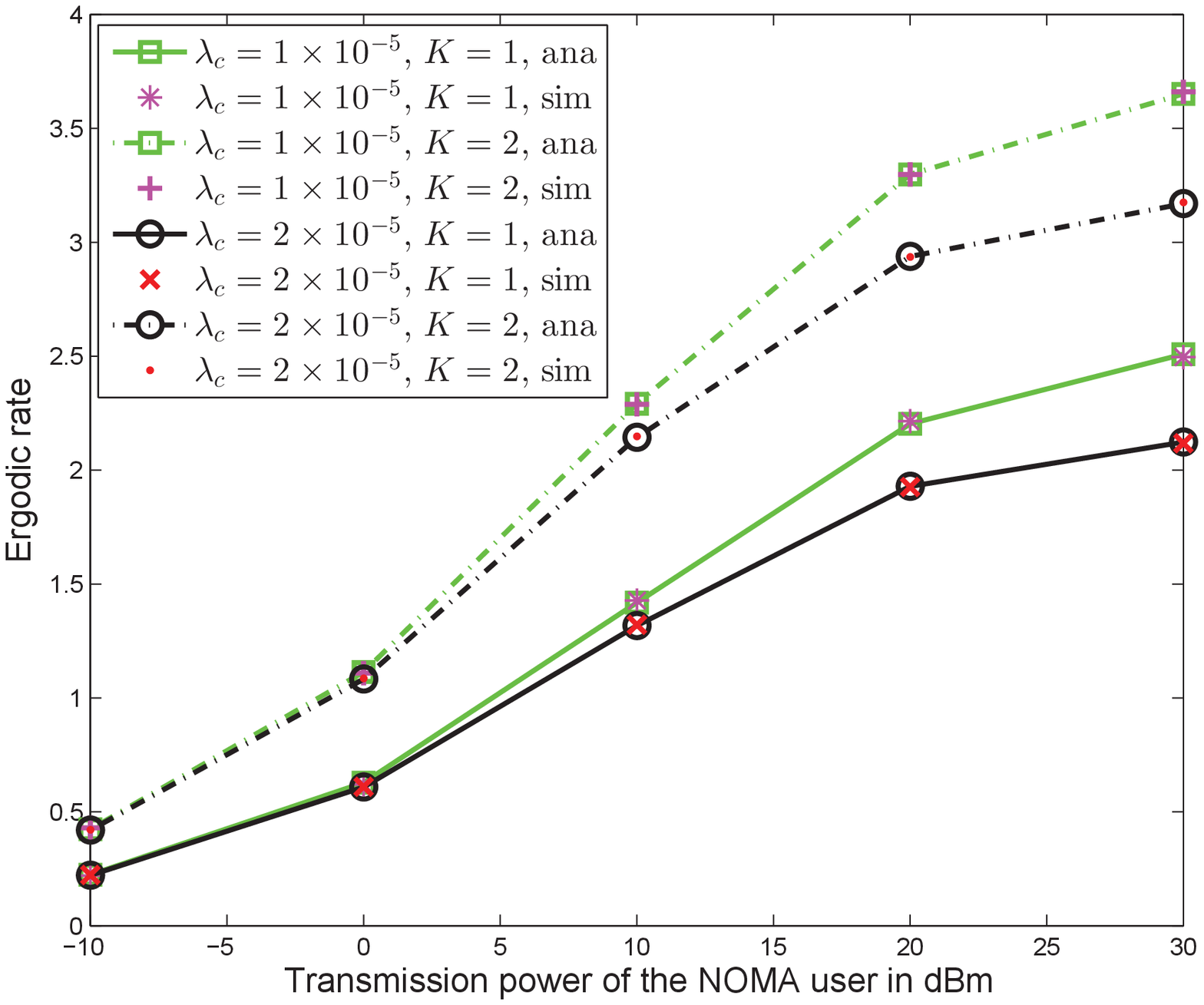}%
\label{NOMA_rate_1}}
\hfil
\subfloat[Erdodic sum rate]{\includegraphics[width=3.2in]{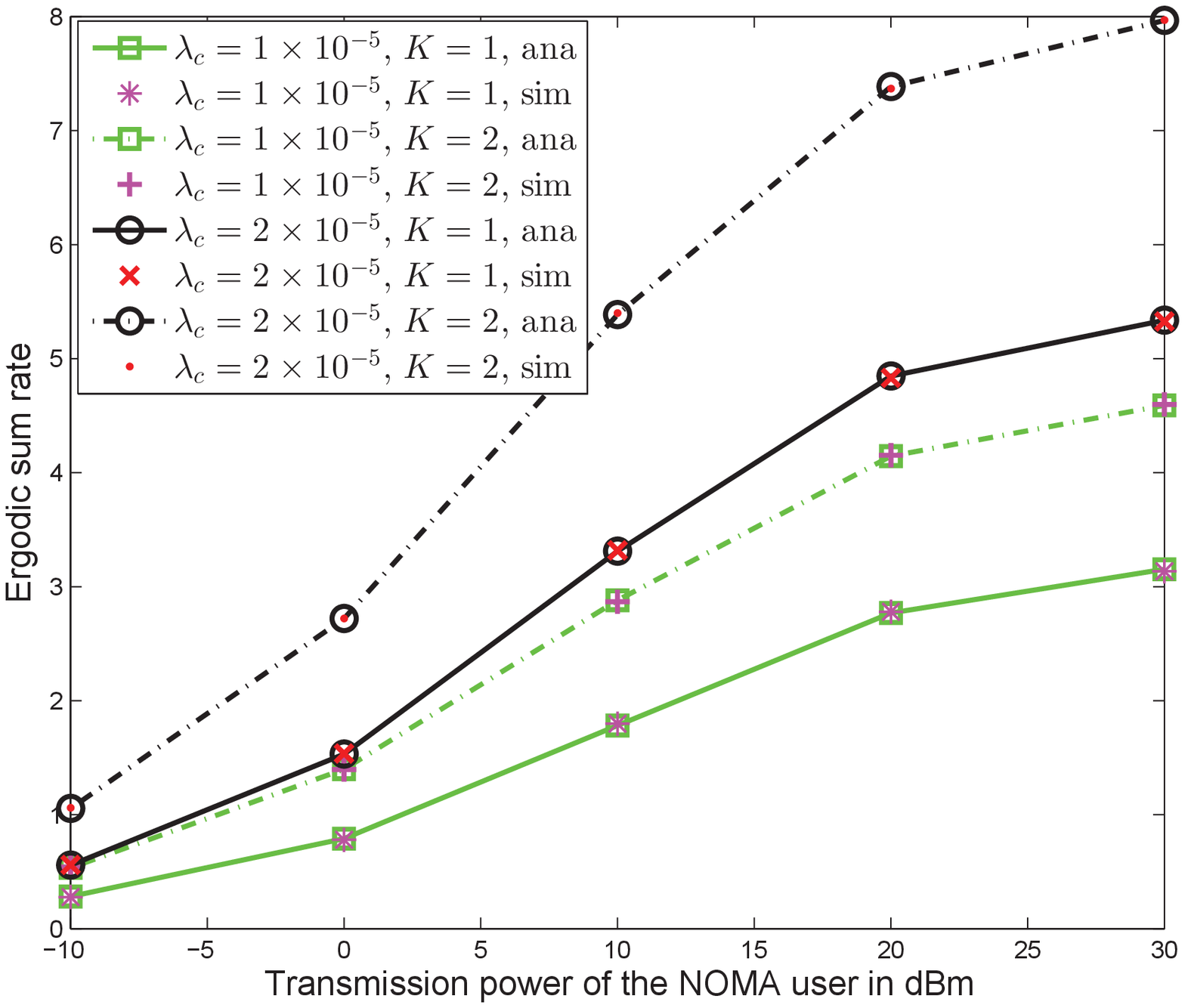}%
\label{NOMA_rate_2}}
\caption{The ergodic rate and {\color{black}ergodic} sum rate of the NOMA users. $\mathcal{R}_c=80$ m, $\phi=10$, and the {\color{black}Gaussian}-Chebyshev parameter $N=20$.}
\label{NOMA_rate}
\end{figure*}

Fig. \ref{NOMA_rate}(a) shows the ergodic rate of a typical NOMA user whose associated base station is located in disk $\mathcal{D}$ and Fig. \ref{NOMA_rate}(b) shows the ergodic sum rate of the NOMA users whose associated base stations are located in disk $\mathcal{D}$. The radius of $\mathcal{D}$ is set as $\mathcal{R}_\mathcal{D}=200$ m in {\color{black}Figs. \ref{NOMA_rate}(a) and \ref{NOMA_rate}(b)}. Both the rate of a {\color{black}typical} NOMA user and the sum rate of the NOMA users increase with $K$, as shown in the figures. {\color{black}This} observation is consistent with the results shown in Fig. \ref{NOMA_outage_1}. It is obvious that as the density of cells increases, the performance of a NOMA user {\color{black}decreases}, due to the {\color{black}enhanced} inter-cell interferences. This {\color{black}statement} is validated in Fig. \ref{NOMA_rate}(a). Interestingly, on the other hand, as the density of cells increases, there will be more NOMA users {\color{black}which} can be served {\color{black}with their} associated base stations located in disk $\mathcal{D}$, which {\color{black}offer} opportunity to improve the sum rate, as shown in Fig. \ref{NOMA_rate}(b).
Thus, there {\color{black}is} a tradeoff between {\color{black}the} single user's QoS and system throughput.

\begin{figure}[!t]
\setlength{\belowcaptionskip}{-1.5em}   
\centering
\includegraphics[width=3.5in]{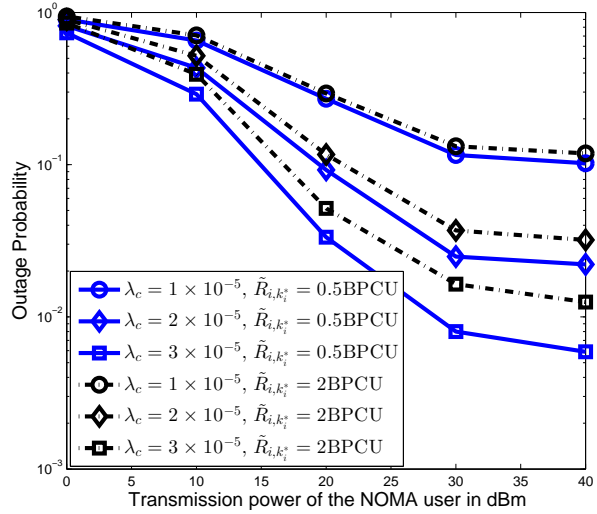}
\caption{{\color{black}CoMP user's outage performance. $\mathcal{R}_\mathcal{D}=300$ m, $\mathcal{R}_c=40$ m, $\phi=20$, $K=3$.}}
\label{CoMP_outage}
\end{figure}

{\color{black}Fig. \ref{CoMP_outage} shows the CoMP user's outage performance, where the NOMA users' data rates are fixed. The CoMP user's data rate is set as $\tilde{R}_0=1$ BPCU. Note that the outage probability achieved by the CoMP user is dependent on the NOMA {\color{black}users'} data rates, because {\color{black}prior to} decoding the CoMP user's message, the base stations need to decode their NOMA {\color{black}users' messages}.
This is consistent with the observation from {\color{black}Fig. \ref{CoMP_outage}}, {\color{black}where} as the rates of NOMA users {\color{black}increase}, the outage performance achieved by the CoMP user degrades. Another interesting observation is that when $\lambda=1\times10^{-5}$, the gap between the two cases with different NOMA users' data rates is fairly small. This can be easily explained, as when $\lambda$ is small, the main limitation of the CoMP user's performance is the number of serving base stations. }

\begin{figure*}[!t]
\setlength{\belowcaptionskip}{-2em}   
\centering
\subfloat[Outage sum rate]{\includegraphics[width=3.2in]{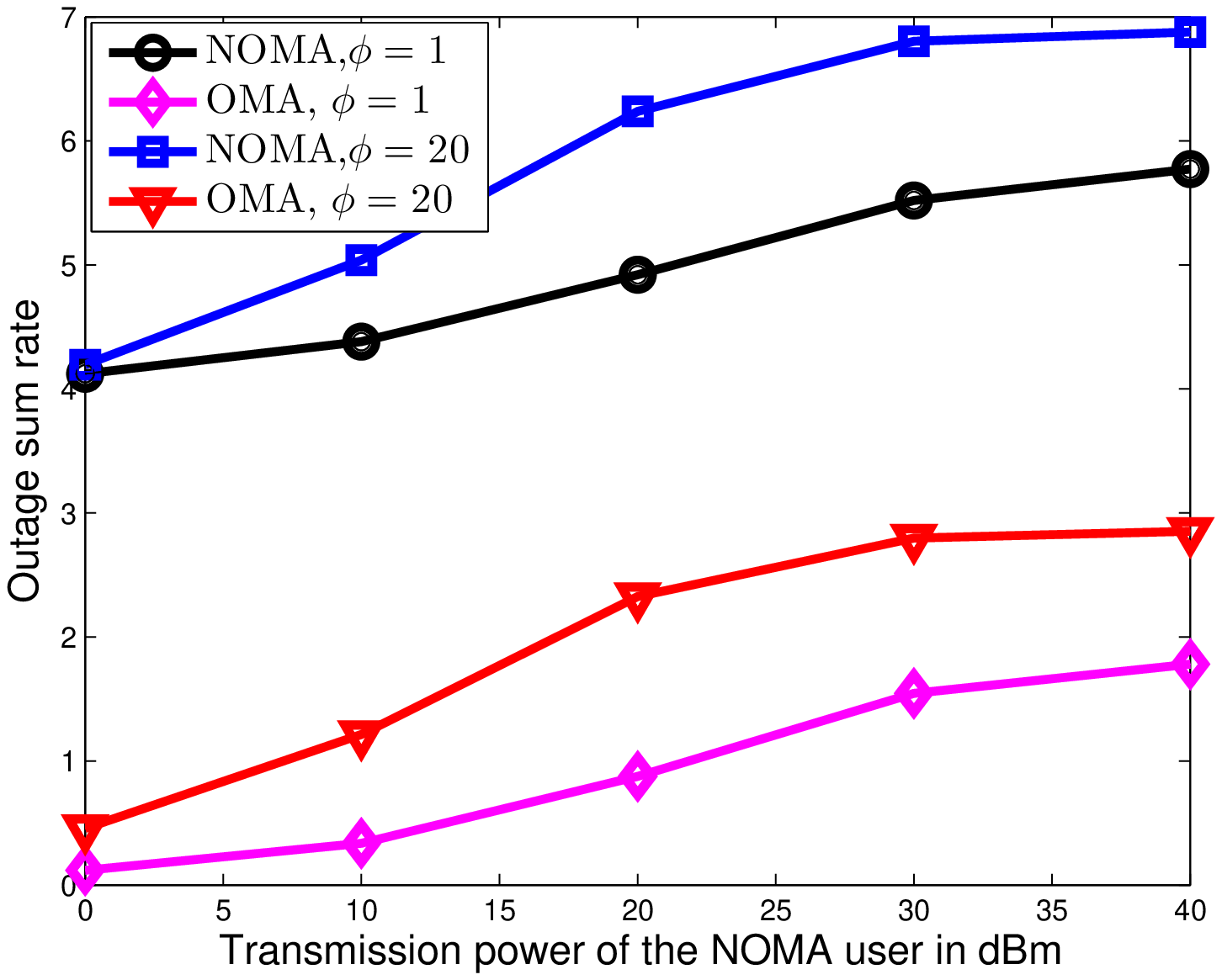}%
\label{NC_Outage sum rate}}
\hfil
\subfloat[Outage probabilities]{\includegraphics[width=3.2in]{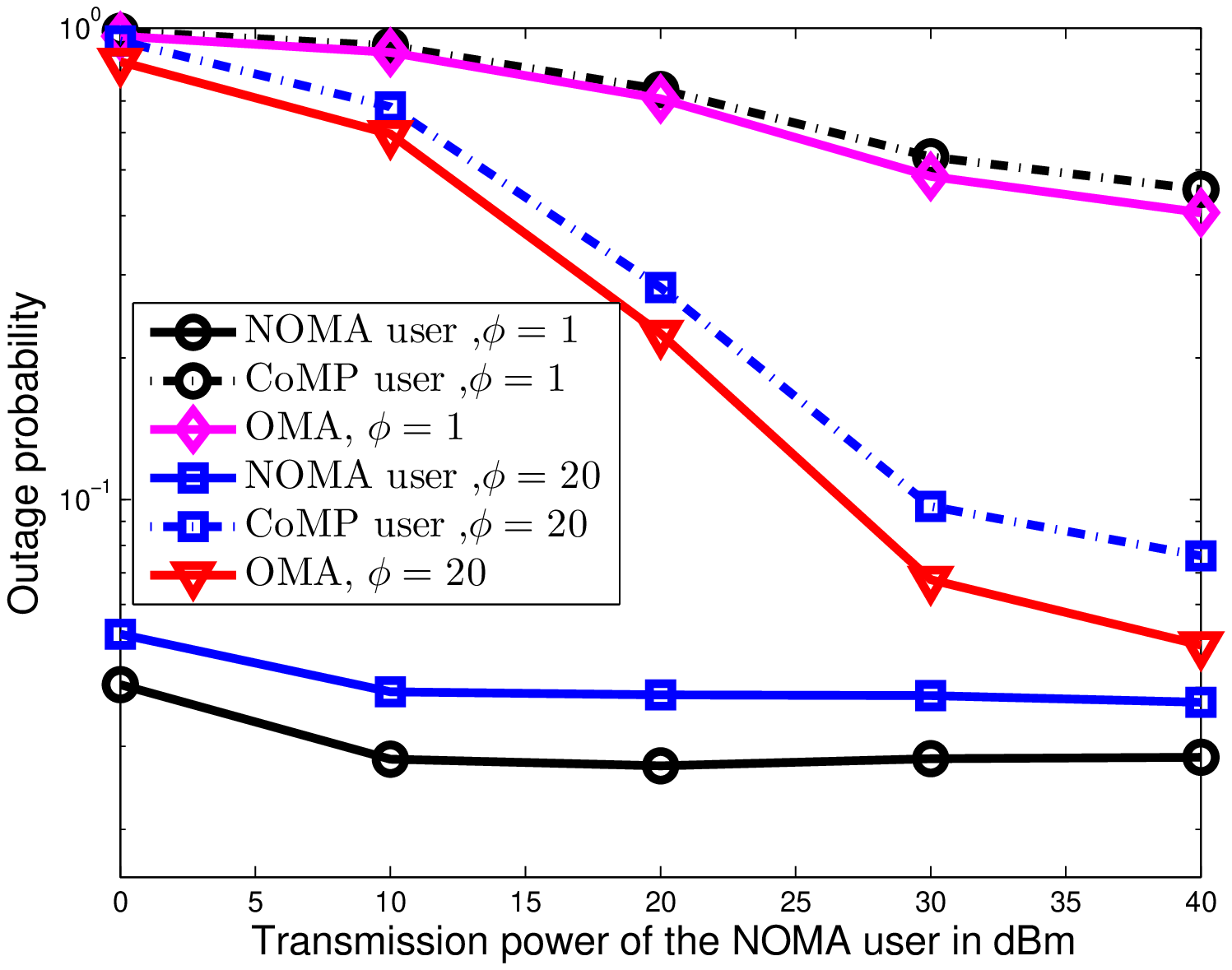}%
\label{NC_Outage probabilities}}
\caption{{\color{black}Performance comparison between N-NOMA and OMA in terms of the outage sum rate. $K=3$, ${\color{black}\tilde{R}_{i,k_i^*}}=0.5$ BPCU, $\tilde{R}_0=3$ BPCU, $\lambda_c=3\times10^{-5}/m^2$, $\mathcal{R}_c=40$ m, $\mathcal{R}_\mathcal{D}=300$ m.
 }}
\label{NC_outagerate}
\end{figure*}

\begin{figure}[!t]
\setlength{\belowcaptionskip}{-1.5em}   
\centering
\includegraphics[width=3.5in]{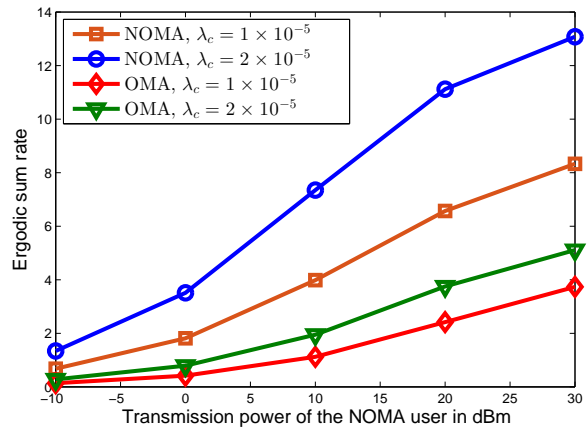}
\caption{{\color{black}Performance comparison between N-NOMA and OMA in terms of the ergodic sum rate. $K=2$, $\phi=10$, $\mathcal{R}_c=80$ m, $\mathcal{R}_\mathcal{D}=200$ m.}}
\label{NC_ergodic}
\end{figure}

{\color{black}Figs. \ref{NC_outagerate} and \ref{NC_ergodic}} show the performance comparison between the proposed N-NOMA scheme and the OMA scheme.
Note that in the benchmark OMA scheme, only the CoMP user is served by the base stations in disk $\mathcal{D}$. {\color{black}More specifically, in the benchmark OMA scheme, if one of the base stations in disk $\mathcal{D}$ can successfully {\color{black}decode} the CoMP user's message through its observation, then the CoMP user is not in outage. }
Fig. \ref{NC_outagerate}(a) shows the performance comparison in terms of the outage sum rate, where the transmission date rates of the users are fixed, {\color{black}while} the corresponding outage probabilities are shown in Fig. \ref{NC_outagerate}(b).
As shown in Fig. \ref{NC_outagerate}(a), the proposed N-NOMA scheme outperforms OMA. For example, when the transmission power of the NOMA user and the CoMP user is $30$ dBm, the outage sum rate achieved by the OMA scheme is about $1.8$ BPCU, {\color{black}while the proposed N-NOMA supports about $5.7$ BPCU. Hence, the N-NOMA scheme has an extra gain of $3.9$ BPCU compared to OMA.} In addition, the proposed N-NOMA scheme also outperforms OMA in terms of the ergodic sum rate, as shown in  Fig. \ref{NC_ergodic}.  {\color{black}Note that, in Fig. \ref{NC_ergodic},  the rates of the NOMA users are adaptively adjusted according to the instantaneous channel conditions as described in Section III. B; thus the CoMP user in the proposed N-NOMA scheme can achieve the same data rates as in the benchmark OMA scheme. {\color{black}Hence, as shown in Fig. \ref{NC_ergodic}, we conclude that the proposed N-NOMA scheme can achieve higher data rates than the benchmark OMA scheme by serving those extra NOMA users, while not degrading the CoMP user's performance.}}

\begin{figure*}[!t]
\setlength{\belowcaptionskip}{-1.5em}   
\centering
\subfloat[]{\includegraphics[width=3.2in]{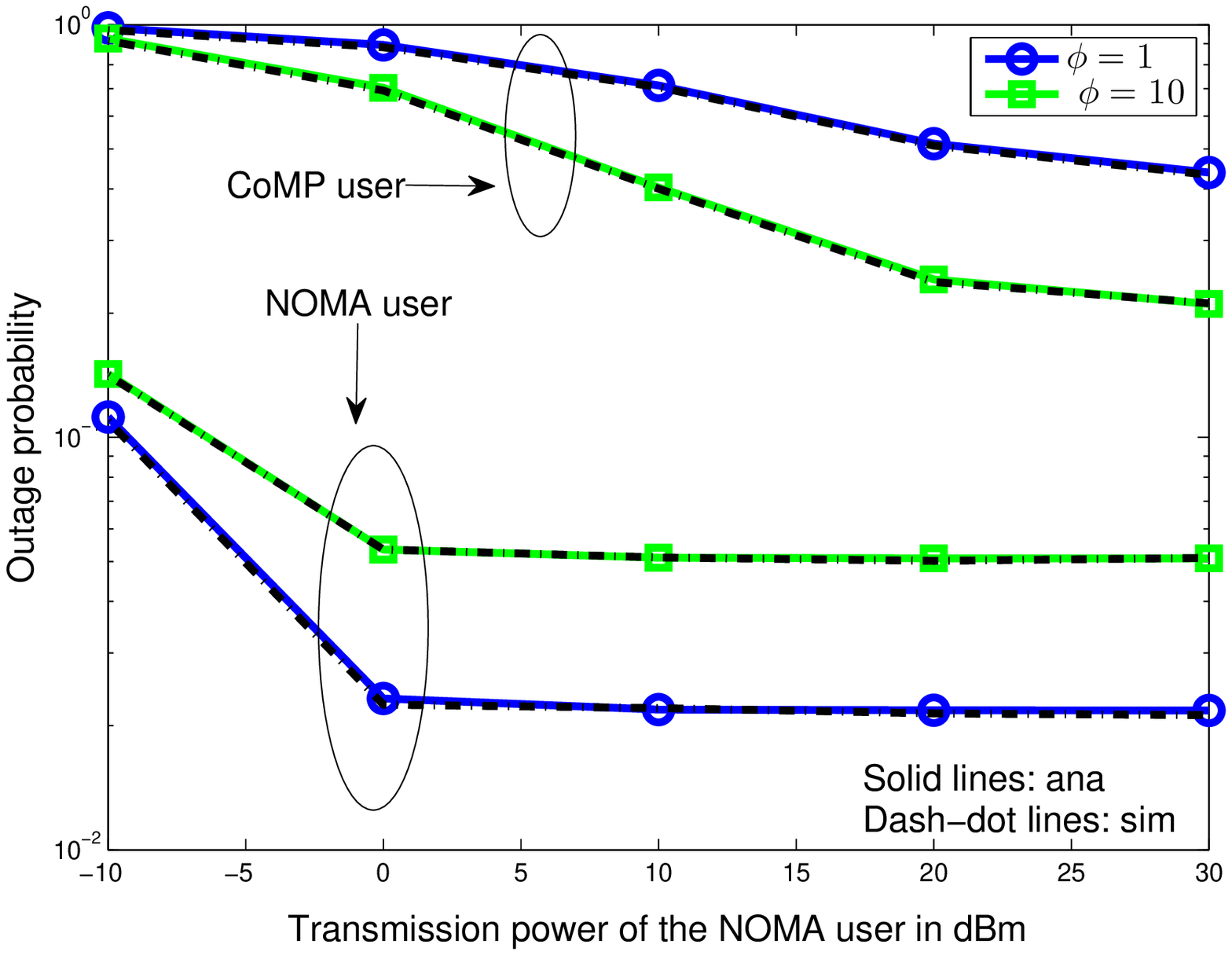}%
\label{neareat_Outage1}}
\hfil
\subfloat[]{\includegraphics[width=3.2in]{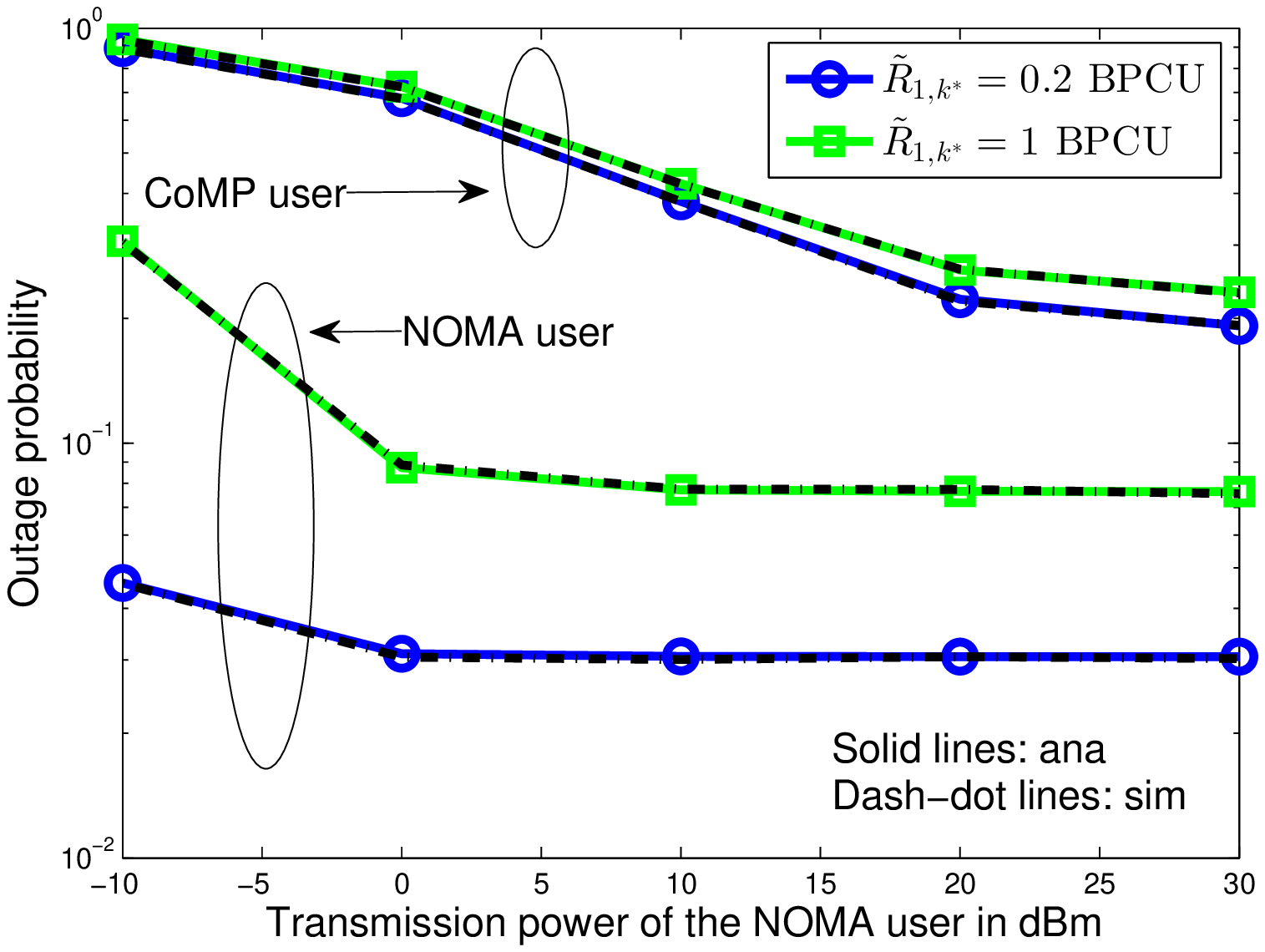}%
\label{neareat_Outage2}}
\caption{{\color{black}Outage probabilities achieved by the nearest N-NOMA scheme. $K=3$, $\mathcal{R}_c=30$ m, $\lambda_c=3\times10^{-5}/m^2$, $\tilde{R}_0=0.5$ BPCU (a) ${\color{black}\tilde{R}_{1,k_1^*}}=0.5$ BPCU (b) $\phi=10$. }}
\label{neareat_Outage}
\end{figure*}
\begin{figure}[!t]
\setlength{\belowcaptionskip}{-1.5em}   
\centering
\includegraphics[width=3.5in]{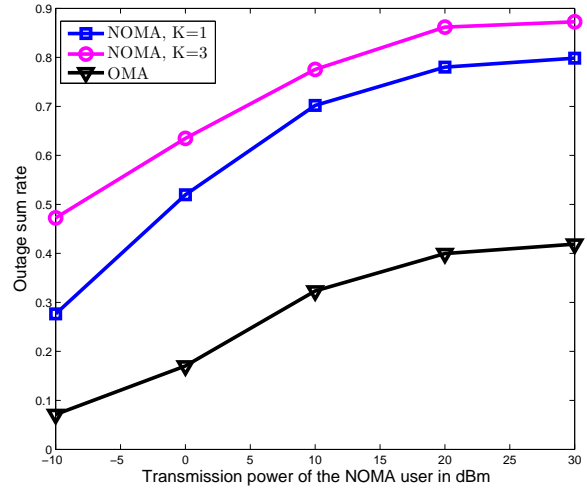}
\caption{{\color{black}Comparison between the nearest N-NOMA scheme and OMA. $\mathcal{R}_c=30$ m, $\lambda_c=3\times10^{-5}/m^2$, $\tilde{R}_0=0.5$ BPCU,  ${\color{black}\tilde{R}_{1,k_1^*}}=0.5$ BPCU, $\phi=10$.}}
\label{neareat_compare}
\end{figure}

{\color{black}Figs. \ref{neareat_Outage} and \ref{neareat_compare}} demonstrate the performance of the nearest N-NOMA scheme. Fig. \ref{neareat_Outage} shows the outage performance achieved by the NOMA {\color{black}and} the CoMP {\color{black}users}, and simulations validate the accuracy of the analysis presented in {\color{black}Proposition} 1 and {\color{black}Proposition} 2.
It is shown in Fig. \ref{neareat_Outage} that in the proposed nearest N-NOMA scheme, the outage probability achieved by the CoMP user is always larger than that {\color{black}of} the NOMA user. This is because only {\color{black}after} the NOMA user's message is decoded, the CoMP user's message {\color{black}can} be decoded.
It is also observed in {\color{black}Figs. \ref{neareat_Outage}(a) and \ref{neareat_Outage}(b)} that increasing the transmission power of the CoMP user {\color{black}and} reducing the date rate of the NOMA user can improve the CoMP user's performance. {\color{black}However}, this is at the expense of degrading the NOMA user's performance.
{\color{black}From Fig. \ref{neareat_Outage}, it can be {\color{black}observed} that the performance of the CoMP user is very poor in the nearest N-NOMA scheme. {\color{black}This} is because only one base station is employed to serve the CoMP user. This observation reveals the importance of inviting multiple base stations to participate in CoMP.}
In Fig. \ref{neareat_compare}, the nearest N-NOMA scheme is compared with the OMA scheme, {\color{black}where the nearest base station is chosen to serve the CoMP user, and only this user is served}. As shown in the figure,  NOMA outperforms OMA in terms of the outage sum rate and the gap between NOMA and OMA {\color{black}increases with $K$}.

\section{Conclusions}
In this paper, in order to {\color{black}improve} {\color{black}the} system throughput and spectral efficiency, we have proposed {\color{black}an} uplink N-NOMA scheme for the CoMP transmission. In the proposed scheme, {\color{black}a} CoMP user and multiple NOMA users are served simultaneously.
To get insight into the problem, a PCP model has been introduced to  characterize the locations of the nodes. {\color{black}The} outage probabilities and ergodic rates of the users have been analyzed to evaluate the performance of the proposed N-NOMA scheme. Closed-form expressions {\color{black}for the outage probabilities and the sum rates} have been developed, {\color{black}and also} validated by computer simulations.
Extensive numerical results have been {\color{black}presented} to demonstrate the performance of the proposed N-NOMA scheme.
Comparisons {\color{black}with} the OMA scheme have been presented, {\color{black}which demonstrate that} N-NOMA can significantly improve the system throughput and spectral efficiency.

\appendices
\section{Proof for Lemma 2}

Recall that $I_{\text{CoMP}}^U  = \frac{\phi|{\color{black}h_{\text{BS}_i,\text{U}_0}}|^2}{ {L\left(||{\color{black}x_i}||\right)}}$. Therefore, the expectation for the CoMP interference can be simply calculated as follows:
\begin{align}
\mathcal{L}_{I_{\text{CoMP}}^U}(s | x_i)&= \mathcal{E}_{I_{\text{CoMP}}^U } \left\{e^{-s I_{\text{CoMP}}^U} \right\}\\\notag
&=\int_0^{\infty} e^{-s\frac{\phi x}{ {L\left(||x_i||\right)}}}e^{-x}\,dx\\\notag
&=\frac{1}{1+s\phi L\left(||x_i||\right)^{-1}},
\end{align}
which follows from the fact that ${\color{black}h_{\text{BS}_i,\text{U}_0}}$ is Rayleigh distributed.

We can evaluate $\mathcal{L}_{I_{\text{inter}}^U}(s|x_i)$ as
\begin{align}\nonumber
\mathcal{L}_{I_{\text{inter}}^U}(s|x_i) &= \mathcal{E} \left\{ \prod_{x_j\in \Phi_c\backslash x_i} \text{exp}\left( -s \frac{|{\color{black}h_{\text{BS}_i,\text{U}_{j,k_j^*}}}|^2}{ {L\left(||{\color{black}y_{j,k_j^*}}+x_j-x_i||\right)}}\right)\right\},
\end{align}
where the arguments for the expectation include ${\color{black}h_{\text{BS}_i,\text{U}_{j,k_j^*}}}$ and the locations of $\text{BS}_i$, $\text{BS}_j$ and $\text{U}_{j,{\color{black}k_j^*}}$.
{\color{black}Note that} ${\color{black}h_{\text{BS}_i,\text{U}_{j,k_j^*}}}$ is independent from $||{\color{black}y_{j,k_j^*}}+x_j-x_i||$. Therefore, by using the assumption that ${\color{black}h_{\text{BS}_i,\text{U}_{j,k_j^*}}}$ is {\color{black}independently and identically} Rayleigh distributed, we have
\begin{align}
\mathcal{L}_{I_{\text{inter}}^U}(s) &= \mathcal{E} \left\{ \prod_{x_j\in \Phi_c\backslash x_i} \frac{1}{ \frac{s}{ {L\left(||{\color{black}y_{j,k_j^*}}+x_j-x_i||\right)}}+1}\right\}.
\end{align}

By applying {\color{black} the Campbell's theorem} and the {\color{black}probability generating functional (PGFL)} \cite{haenggi2012stochastic}, we have
\begin{align}\label{lap_inter}
\mathcal{L}_{I_{\text{inter}}^U}(s) =& \text{exp}\left(-\lambda_{c} \int_{R^2} \left( 1 -\mathcal{E}_{{\color{black}y_{j,k_j^*}}}\left\{\frac{1}{ \frac{s}{ {L\left(||{\color{black}y_{j,k_j^*}}+x-x_i||\right)}}+1}\right\} \right) dx \right).
\end{align}
Denote the pdf of ${\color{black}y_{j,k_j^*}}$ by $f_{{\color{black}y_{j,k_j^*}}} (y)$, and the Laplace transform is expressed as follows:
\begin{align}
\mathcal{L}_{I_{\text{inter}}^U}(s) =& \text{exp}\left(-\lambda_{c} \int_{R^2}f_{{\color{black}y_{j,k_j^*}}} (y) \int_{R^2} \left(1 - \frac{1}{ \frac{s}{ {L\left(||y +x-x_i||\right)}}+1} \right) dx dy\right).
\end{align}
Using the  {\color{black}substitution} $y+x-x_i\rightarrow x'$, we have
\begin{align}
\mathcal{L}_{I_{\text{inter}}^U}(s) =& \text{exp}\left(-\lambda_{c} \int_{R^2}f_{{\color{black}y_{j,k_j^*}}} (y) \int_{R^2} \left( 1- \frac{1}{ \frac{s}{ {L\left(||x'||\right)}}+1}  \right) dx'dy \right)\\ \nonumber
\overset{(a)}{=}& \text{exp}\left(-\lambda_{c} \int_{R^2}f_{{\color{black}y_{j,k_j^*}}} (y) 2\pi \int_0^\infty  \left( 1- \frac{1}{ \frac{s}{ {L\left(r\right)}}+1} \right) rdrdy \right),
\end{align}
where $(a)$ follows from changing to polar coordinates.

{\color{black}Further, by applying} {\color{black}the} Beta function, we have
 \begin{align}\nonumber
\mathcal{L}_{I_{\text{inter}}^U}(s) \overset{(a)}{=}& \text{exp}\left(-\lambda_{c} \int_{R^2}f_{{\color{black}y_{j,k_j^*}}} (y) 2\pi \frac{s^{\frac{2}{\alpha}}}{\alpha}\text{B}\left(\frac{2}{\alpha}, \frac{\alpha-2}{\alpha}\right)dy \right)
\\ \overset{(b)}{=}&
\text{exp}\left(-   2\pi\lambda_{c} \frac{s^{\frac{2}{\alpha}}}{\alpha}\text{B}\left(\frac{2}{\alpha}, \frac{\alpha-2}{\alpha}\right) \right),
\end{align}
{\color{black}where $(a)$ follows from the fact that the integral with respect to $r$ is not a function of $y$, and $(b)$ follows from the fact that $\int_{R^2}f_{{\color{black}y_{j,k_j^*}}} (y)\,dy=1$}. The use of other user selection {\color{black}strategies} does not affect much the results for the {\color{black}Laplace of the interference}.  Please note that $\alpha$ cannot be chosen to be $2$.

Therefore Lemma 2 is proved.

\section{Proof for Theorem 1}
The SINR to decode $\text{U}_{i,k_i^*}$'s message can be expressed as follows:
\begin{align}
 {\color{black}\text{SINR}_{k_i^*}^{\text{BS}_i}} = \frac{{\color{black}z_{k_i^*,i}}}{  I_{\text{CoMP}}^U   +I_{\text{inter}}^U
 +\frac{1}{\rho}}.
\end{align}
Since the Laplace transform for $I_{\text{CoMP}}^U$ is a function of $||x_i||$, we first fix the location of $\text{BS}_i$ and calculate the conditioned outage probability:
\begin{align}
 {\color{black}\mathrm{P}_{\text{NOMA}}} (x_i)
 =&\mathrm{P}\left(\text{SINR}_{{\color{black}k_i^*}}^{\text{BS}_i}<\epsilon_{i} |x_i\right)\\\notag
 =& \mathrm{P}\left( z_{{\color{black}k_i^*},i}<\epsilon_i\left( I_{\text{CoMP}}^U  +I_{\text{inter}}^U
 +\frac{1}{\rho}\right)\right)\\ \nonumber
 =& \mathcal{E}_{I_{\text{CoMP}}^U ,I_{\text{inter}}^U }\left\{\left(\mathrm{P}\left( z_{k,i}<\epsilon_i\left( I_{\text{CoMP}}^U  +I_{\text{inter}}^U
 +\frac{1}{\rho}\right)\right)\right)^K\right\},
\end{align}
where the last step follows from {\color{black}the fact} that $z_{{\color{black}k_i^*},i}$ is largest in  $\{z_{1,i}, \ldots,  z_{K,i}\}$.
By applying the approximated expression for the pdf of $z_{k,i}$ as shown in {\color{black}Lemma} 1, we have
\begin{align}
 &{\color{black}\mathrm{P}_{\text{NOMA}}}(x_i)\\\notag
 &\approx \mathcal{E}_{I_{\text{CoMP}}^U ,I_{\text{inter}}^U }\left\{ \left(\sum^{N}_{n=1}w_n  \left(1-e^{-c_n\epsilon_i\left( I_{\text{CoMP}}^U  +I_{\text{inter}}^U
 +\frac{1}{\rho}\right)}\right)\right)^K \right\}
 \\\notag
 &= \mathcal{E}_{I_{\text{CoMP}}^U ,I_{\text{inter}}^U }\left\{ \left(\sum^{N}_{n=1}w_n-\sum^{N}_{n=1}w_n  e^{-c_n\epsilon_i\left( I_{\text{CoMP}}^U  +I_{\text{inter}}^U
 +\frac{1}{\rho}\right)}\right)^K \right\}
 \\\nonumber
 &{\approx} \mathcal{E}_{I_{\text{CoMP}}^U ,I_{\text{inter}}^U }\left\{  \left(1-\sum^{N}_{n=1}w_n e^{-c_n\epsilon_i\left( I_{\text{CoMP}}^U  +I_{\text{inter}}^U
 +\frac{1}{\rho}\right)}\right)^K \right\}.
 \end{align}
 Define $\tilde{w}_n=-w_n$ for $1\leq n \leq N$ and $\tilde{w}_0=1$. In addition, $\tilde{c}_n=c_n$ for $1\leq n \leq N$ and $\tilde{c}_0=0$.
 Applying the  multinomial theorem, we expand the expression as follows:
 \begin{align}\nonumber
 {\color{black}\mathrm{P}_{\text{NOMA}}}(x_i) \approx& 1+\sum_{\substack{k_0+\cdots+k_N=K \\ k_0\neq K}}{K \choose k_0, \cdots, k_N} \prod_{n=0,k_n\neq 0}^{N} \tilde{w}_n^{k_n} e^{-k_n\tilde{c}_n\epsilon_i
 \frac{1}{\rho}}  \\   &\times    \mathcal{E}_{I_{\text{CoMP}}^U   } \left\{e^{-\mu\epsilon_i I_{\text{CoMP}}^U  } |x_i\right\}   \mathcal{E}_{I_{\text{inter}}^U } \left\{e^{-\mu\epsilon_i  I_{\text{inter}}^U} |x_i\right\},
\end{align}
where $\mu= \sum_{n=0,k_n\neq 0}^{N}k_n\tilde{c}_n$, ${K \choose k_0, \cdots,k_N}=\frac{K!}{k_0!\cdots k_N!}$.

By applying the Laplace transform for $I_{\text{CoMP}}^U$ and $I_{\text{Inter}}^U$ obtained in {\color{black}Lemma 2}, the conditioned outage  probability can be expressed as
\begin{align}
 {\color{black}\mathrm{P}_{\text{NOMA}}}(x_i) \approx&1+\sum_{\substack{k_0+\cdots+k_N=K \\ k_0\neq K}}{K \choose k_0, \cdots, k_N} \prod_{n=0,k_n\neq 0}^{N} \tilde{w}_n^{k_n} e^{-k_n\tilde{c}_n\epsilon_i
 \frac{1}{\rho}}  \\ \nonumber &\times    \frac{
\text{exp}\left(-   2\pi\lambda_{c} \frac{\mu^{\frac{2}{\alpha}}}{\alpha}\text{B}\left(\frac{2}{\alpha}, \frac{\alpha-2}{\alpha}\right) \right)}{1+\phi\mu L\left(||x_i||\right)^{-1}}.
\end{align}

Note that $x_i$ is uniformly distributed in disk $\mathcal{D}$, {\color{black}for the reason that we have conditioned on $x_i \in \Phi_c \cap \mathcal{D}$}. Thus,  the outage probability ${\color{black}\mathrm{P}_{\text{NOMA}}}$ can be calculated as follows:
\begin{align}
 {\color{black}\mathrm{P}_{\text{NOMA}}}=&\frac{1}{\pi\mathcal{R_{\mathcal{D}}}^2}\int_{x_i \in \mathcal{D}}{\color{black}\mathrm{P}_{\text{NOMA}}}(x_i)\,dx_i\\\notag
\overset{(a)}{=}&1+\frac{2}{\mathcal{R_{\mathcal{D}}}^2}\sum_{\substack{k_0+\cdots+k_N=K \\ k_0\neq K}}{K \choose k_0, \cdots, k_N} \prod_{n=0,k_n\neq 0}^{N} \tilde{w}_n^{k_n}\\\notag
 &\times e^{-k_n\tilde{c}_n\epsilon_i
 \frac{1}{\rho}} {
\text{exp}\left(-   2\pi\lambda_{c} \frac{\mu^{\frac{2}{\alpha}}}{\alpha}\text{B}\left(\frac{2}{\alpha}, \frac{\alpha-2}{\alpha}\right) \right)}
\\ \nonumber &\times   \int_0^{R_{\mathcal{D}}}   \frac{  1 }{1+\phi\mu r^{-\alpha}} rdr,
\end{align}
where (a) follows from changing to polar coordinates.
Finally, by applying  {\color{black}the} hypergeometric function, {\color{black}Theorem 1} is proved.

\section{Proof for Corollary 1}
\begin{align}
{\color{black}R_{\text{NOMA}}^{ave}}&=\mathcal{E}\left\{\log_2(1+\text{SINR}_{{\color{black}k_i^*}}^{\text{BS}_i})\right\}\\\notag
&\overset{(a)}{=}\int_0^{\infty}\mathrm{P}\left(\log_2(1+\text{SINR}_{{\color{black}k_i^*}}^{\text{BS}_i}) > t\right)\,dt\\\notag
&=\int_0^{\infty}\mathrm{P}\left(\text{SINR}_{{\color{black}k_i^*}}^{\text{BS}_i} > 2^t-1\right)\,dt\\\notag
&=\int_0^{\infty}1-\mathrm{P}\left(\text{SINR}_{{\color{black}k_i^*}}^{\text{BS}_i} < 2^t-1\right)\,dt\\\notag
&\overset{(b)}{=}\frac{1}{\ln{2}}\int_0^{\infty}\frac{1-\mathrm{P}\left(\text{SINR}_{{\color{black}k_i^*}}^{\text{BS}_i} < x\right)}{1+x}\,dx,
\end{align}
where $(a)$ is {\color{black}obtained} as $\mathcal{E}\{x\}=\int_{t>0}\mathrm{P}(X>t)\,dt$ for a positive random variable X and it is obvious that $\log_2(1+\text{SINR}_{{\color{black}k_i^*}}^{\text{BS}_i})$ is a positive random variable, and $(b)$ follows from $x=2^t-1$.

After some manipulations, {\color{black}Corollary 1} is obtained.
\section{}
\subsection{Proof for {\color{black}Proposition 1}}
Denote the distance between the CoMP user to its nearest BS as $d$, i.e., $d=L(||{\color{black}x_1}||)$. We first consider the case when the value of $d$ is fixed.

Given a fixed $d$, the Laplace transform of interference from {\color{black}the} other NOMA users can be expressed as
\begin{align}
\mathcal{L}_{I_{\text{inter}}^U}(s) &= \mathcal{E} \left\{ \prod_{x_j\in \Phi_c\backslash x_1} \frac{1}{ \frac{s}{ {L\left(||{\color{black}y_{j,k_j^*}}+x_j-x_1||\right)}}+1}\right\}.
\end{align}

Again, by applying {\color{black}the Campbell's theorem} and the PGFL \cite{haenggi2012stochastic}, {\color{black}the following expression} is obtained
\begin{align}\label{lap_nearest}
\mathcal{L}_{I_{\text{inter}}^U}(s) =& \text{exp}\left(-\lambda_{c} \int_{||x||>d^2} \left( 1 \right.\right. \\\nonumber &\left.\left.-\mathcal{E}_{{\color{black}y_{j,k_j^*}}}\left\{\frac{1}{ \frac{s}{ {L\left(||{\color{black}y_{j,k_j^*}}+x-x_1||\right)}}+1}\right\} \right) dx \right).
\end{align}
{\color{black}Note that in (\ref{lap_nearest}), the integral region is changed compared to (\ref{lap_inter}), which makes the calculation more challenging, because the effect of ${\color{black}y_{j,k_j^*}}$ cannot be removed from (\ref{lap_nearest}) as in (\ref{lap_inter})}.

A closed-form expression for $\mathcal{E}_{{\color{black}y_{j,k_j^*}}}\left\{\frac{1}{ \frac{s}{ {L\left(||{\color{black}y_{j,k_j^*}}+x-x_1||\right)}}+1}\right\}$, when $K=1$, can be obtained using a geometric probability approach as in \cite{tabassum2014interference,hina2017uplinkNOMAPCP}. However, for $K>1$, the calculation of $\mathcal{E}_{{\color{black}y_{j,k_j^*}}}\left\{\frac{1}{ \frac{s}{ {L\left(||{\color{black}y_{j,k_j^*}}+x-x_1||\right)}}+1}\right\}$ {\color{black}is} very challenging.
{\color{black}Hence}, making  reasonable {\color{black}approximations} to simplify the calculation is  necessary. Note that when the distance between the BSs is much {\color{black}larger} than that between a NOMA user {\color{black}and} its associated BS, the impact of ${\color{black}y_{j,k_j^*}}$ can be omitted. Taking $\alpha=4$, $\mathcal{L}_{I_{\text{inter}}^U}(s)$ can be approximated as:
 \begin{align}\label{Lap1}
\mathcal{L}_{I_{\text{inter}}^U}(s)    \approx&
\text{exp}\left(-2\pi\lambda_{c} \frac{s^{\frac{2}{\alpha}}}{\alpha}\text{B}\left(\frac{2}{\alpha}, \frac{\alpha-2}{\alpha}\right)\right.\\
&\left.+\frac{1}{2}\pi\lambda_c\sqrt{s}
\arctan\left(\sqrt{-\frac{1}{2}+\frac{1}{2}\sqrt{1+\frac{16d^4}{s}}}\right)\right).
\end{align}
Then, by using  similar steps as in Appendix B, the {\color{black}outage} probability of the nearest BS's NOMA user given a fixed $d$ and fixed $|{\color{black}h_{\text{BS}_i,\text{U}_0}}|$ can be approximated as {\color{black}follows:}
\begin{align}
 &\mathrm{P}_{1}^{\text{nearest}}(d,|{\color{black}h_{\text{BS}_i,\text{U}_0}}|)\\\notag
 & \approx1+ \sum_{\substack{k_0+\cdots+k_N=K \\ k_0\neq K}}{K \choose k_0, \cdots, k_N} \prod_{n=0,k_n\neq 0}^{N} \tilde{w}_n^{k_n} e^{-k_n\tilde{c}_n\epsilon_1
 \frac{1}{\rho}}  \\ \nonumber &\times
\text{exp}\left(-2\pi\lambda_{c} \frac{(\epsilon_1\mu)^{\frac{2}{\alpha}}}{\alpha}\text{B}\left(\frac{2}{\alpha}, \frac{\alpha-2}{\alpha}\right)\right.\\ \notag
&\left.\quad\quad+\frac{1}{2}\pi\lambda_c\sqrt{\epsilon_1\mu}
\arctan\left(\sqrt{-\frac{1}{2}+\frac{1}{2}\sqrt{1+\frac{16d^4}{\epsilon_1\mu}}}\right)
\right)\\\notag
&\times \text{exp}\left(-\frac{\phi\epsilon_1\mu|{\color{black}h_{\text{BS}_i,\text{U}_0}}|^2}{d^4}\right)  .
\end{align}
After taking {\color{black}the} average over ${\color{black}h_{\text{BS}_i,\text{U}_0}}$, the {\color{black}outage} probability of the nearest BS's NOMA user given a fixed $d$ can be approximated as:
\begin{align}
 \mathrm{P}_{1}^{\text{nearest}}(d) \approx&1+ \sum_{\substack{k_0+\cdots+k_N=K \\ k_0\neq K}}{K \choose k_0, \cdots, k_N} \prod_{n=0,k_n\neq 0}^{N} \tilde{w}_n^{k_n} e^{-k_n\tilde{c}_n\epsilon_1
 \frac{1}{\rho}}  \\ \nonumber &\times
\text{exp}\left(-2\pi\lambda_{c} \frac{(\epsilon_1\mu)^{\frac{2}{\alpha}}}{\alpha}\text{B}\left(\frac{2}{\alpha}, \frac{\alpha-2}{\alpha}\right)\right.\\ \notag
&\left.\quad\quad+\frac{1}{2}\pi\lambda_c\sqrt{\epsilon_1\mu}
\arctan\left(\sqrt{-\frac{1}{2}+\frac{1}{2}\sqrt{1+\frac{16d^4}{\epsilon_1\mu}}}\right)
\right)\\\notag
&\times \frac{1}{1+\phi\epsilon_1\mu d^{-4}}  .
\end{align}

Note that the pdf of $L(||{\color{black}x_1}||)$ \cite{haenggi2012stochastic} can be expressed as
\begin{align}
f_{L(||{\color{black}x_1}||)}(d)=2\lambda_c\pi d e^{ -\lambda_c\pi d^2}.
\end{align}

After taking {\color{black}the} average over $L(||{\color{black}x_1}||)$, the expression in {\color{black}Proposition 1} can be {\color{black}easily} obtained.
{\color{black}\subsection{Proof for Proposition 2}
To successfully decode CoMP user's message, the following two conditions must be met.
\begin{enumerate}
  \item The NOMA user's message can be successfully decoded, i.e., $\text{SINR}^{\text{BS}_1}_{k_1^*}>\epsilon_1$
  \item After removing the NOMA user's message, the CoMP user's message can be decoded, i.e.,  $\frac{\frac{\phi|{\color{black}h_{\text{BS}_1,\text{U}_0}}|^2}{d^4}}{I_{\text{inter}}^{U}+\frac{1}{\rho}}>\epsilon_0$.
\end{enumerate}
{\color{black}As such}, the outage probability of the CoMP user can be expressed as
\begin{align}
\mathrm{P}_0^{\text{nearest}}&=\mathcal{E}_{d,I^U_{\text{inter}}}\left\{\mathrm{P}_0^{\text{nearest}}\left(d,I^U_{\text{inter}}\right)\right\},
\end{align}
where $\mathrm{P}_0^{\text{nearest}}\left(d,I^U_{\text{inter}}\right)$ is the outage probability given a fixed $d$ and $I^U_{\text{inter}}$, and can be calculated as follows:
\begin{align}
\mathrm{P}_0^{\text{nearest}}\left(d,I^U_{\text{inter}}\right)
&=1-\int_{\frac{\epsilon_0d^4(I^U_{\text{inter}}+1/\rho)}{\phi}}^{\infty}
\int_{\epsilon_1\left(\frac{\phi x}{d^4}+I^U_{\text{inter}}+\frac{1}{\rho}\right)}^{\infty}
f_{|h_{\text{BS}_1,\text{U}_0}|^2}(x)f_{z_{k_1^*,1}}(y)\,dxdy\\\notag
&=1+ \sum_{\substack{k_0+\cdots+k_N=K \\ k_0\neq K}}{K \choose k_0, \cdots, k_N} \prod_{n=0,k_n\neq 0}^{N} \tilde{w}_n^{k_n} e^{-k_n\tilde{c}_n\epsilon_1
 \frac{1}{\rho}}  \\ \nonumber &\times
\text{exp}\left(-\left(\epsilon_1\mu+\frac{\epsilon_0(\phi\epsilon_1\mu+d^4)}{\phi}\right)I^U_{\text{inter}}\right)\\\notag
&\times \frac{d^4}{\phi\epsilon_1\mu+d^4}\text{exp}\left(-\frac{\epsilon_0(\phi\epsilon_1\mu+d^4)}{\phi\rho}\right).
\end{align}
By applying the Laplace transform of $I^U_{\text{inter}}$ as expressed in (\ref{Lap1}) and taking {\color{black}the} average over $d$, the expression in (\ref{P_0_n}) is obtained and the proof for {\color{black}Proposition 2} is  {\color{black}complete}.}
\bibliographystyle{IEEEtran}
\bibliography{IEEEabrv,ref}
\end{document}